\documentclass[aps,pre,reprint,amsmath,amssymb,superscriptaddress,showpacs,floatfix]{revtex4-1}
\usepackage{graphicx}
\usepackage{dcolumn}
\usepackage{bm}
\usepackage[colorlinks, allcolors=blue]{hyperref}
\usepackage[usenames, dvipsnames]{color}

\begin{document}

\title{Low temperature thermodynamics of the antiferromagnetic $J_1-J_2$ model: Entropy, critical points and spin gap}

 \author{Sudip Kumar Saha}
% \email{sud.dream@gmail.com}
 \affiliation{S. N. Bose National Centre for Basic Sciences, Block - JD, Sector - III, Salt Lake, Kolkata - 700106, India}

 \author{Manodip Routh}
 \affiliation{S. N. Bose National Centre for Basic Sciences, Block - JD, Sector - III, Salt Lake, Kolkata - 700106, India}

\author{Manoranjan Kumar}
 \email{manoranjan.kumar@bose.res.in}
 \affiliation{S. N. Bose National Centre for Basic Sciences, Block - JD, Sector - III, Salt Lake, Kolkata - 700106, India}

 \author{Zolt\'an G. Soos}
 \email{soos@princeton.edu}
 \affiliation{Department of Chemistry, Princeton University, Princeton, New Jersey 08544, USA}

\date{\today}

\begin{abstract}
The antiferromagnetic $J_1-J_2$ model is a spin-1/2 chain with isotropic exchange $J_1 > 0$ between first neighbors and $J_2 = \alpha J_1$ 
between second neighbors. The model supports both gapless quantum phases with nondegenerate ground states and gapped phases 
with $\Delta(\alpha) > 0$ and doubly degenerate ground states. Exact thermodynamics is limited to $\alpha = 0$, 
the linear Heisenberg antiferromagnet (HAF). Exact diagonalization of small systems at frustration $\alpha$ followed by 
density matrix renormalization group (DMRG) calculations returns the entropy density $S(T,\alpha,N)$ and magnetic susceptibility $\chi(T,\alpha,N)$ 
of progressively larger systems up to $N = 96$ or 152 spins. Convergence to the thermodynamics limit, $S(T,\alpha)$ or $\chi(T,\alpha)$, 
is demonstrated down to $T/J \sim 0.01$ in the sectors $\alpha < 1$ and $\alpha > 1$. $S(T,\alpha)$ yields the critical points between gapless 
phases with $S^\prime(0,\alpha) > 0$ and gapped phases with $S^\prime(0,\alpha) = 0$. The $S^\prime(T,\alpha)$ maximum at $T^*(\alpha)$ 
is obtained directly in chains with large $\Delta(\alpha)$ and by extrapolation for small gaps. A phenomenological approximation for $S(T,\alpha)$ 
down to $T = 0$ indicates power-law deviations $T^{-\gamma(\alpha)}$ from $\exp(-\Delta(\alpha)/T)$ with exponent $\gamma(\alpha)$ that increases with $\alpha$. 
The $\chi(T,\alpha)$ analysis also yields power-law deviations, but with exponent $\eta(\alpha)$ that decreases with $\alpha$. 
%Spin correlation functions account for $S(T,\alpha)$ differences between frustration $\alpha < 1$ within a chain and $\alpha > 1$ between HAFs on sublattices. 
$S(T,\alpha)$ and the spin density $\rho(T,\alpha) = 4T\chi(T,\alpha)$ probe the thermal and magnetic fluctuations, respectively, 
of strongly correlated spin states. Gapless chains have constant  $S(T,\alpha)/\rho(T,\alpha)$ for $T < 0.10$. Remarkably, the ratio decreases 
(increases) with $T$ in chains with large (small) $\Delta(\alpha)$.

\end{abstract}

\maketitle
%%%%%%%%%%%%%%%%%%%%%%%%%%%%%%%%%%%%%%%%%%%%%%%%%%%%%%%%%%%%%%%%%%
%																 %
%		INTRODUCTION											 %
%																 %
%%%%%%%%%%%%%%%%%%%%%%%%%%%%%%%%%%%%%%%%%%%%%%%%%%%%%%%%%%%%%%%%%%

\section{\label{sec1} Introduction}

The antiferromagnetic $J_1-J_2$ model, Eq.~\ref{eq:j1j2} below, is in the large family of 1D models with one spin per unit cell that 
includes Heisenberg, Ising, XY and XXZ chains, among others. Their rich quantum ($T = 0$) phase diagrams have fascinated theorists 
for decades in such contexts as field theory, critical phenomena, density matrix renormalization group (DMRG) calculations, exact 
many-spin results and the unexpected difference between spin-1/2 and spin-1 Heisenberg chains. A uniform magnetic field $B$ and 
ferromagnetic exchange expand the variety of exotic quantum phases. 

Exact thermodynamics, aside from some Ising models, is limited to the linear Heisenberg antiferromagnet~\cite{johnstonprl} (HAF). 
Maeshima and Okunishi~\cite{okunishiprb} studied the thermodynamics of Eq.~\ref{eq:j1j2} at both $B = 0$ and $B > 0$ using the 
transfer matrix renormalization group (TMRG). Feigun and White~\cite{white2005} obtained the $B = 0$ thermodynamics with an enlarged 
Hilbert space with ancilla. The methods agree quantitatively for $T/J > 0.2$ and semi-quantitatively down to $T/J \sim 0.1$. 
In this paper, we discuss the thermodynamics of Eq.~\ref{eq:j1j2} using exact diagonalization (ED) of short chains followed by DMRG 
calculations of the low-energy states of progressively longer chains in which the thermodynamic limit holds down to progressively 
lower $T$. We lower the converged range to $T/J \sim 0.01$. Exact HAF thermodynamics~\cite{johnstonprl} reaches decades lower $T$ where logarithmic contributions are important.

The antiferromagnetic $J_1-J_2$ model is a spin-1/2 chain with isotropic exchange $J_1$ and $J_2$ between first and second neighbors, 
respectively. The model at frustration $\alpha = J_2/J_1$ is conventionally written with $J_1 = 1$ as
\begin{equation}
H(\alpha) = \sum_{r} \vec{S}_r \cdot \vec{S}_{r+1} + \alpha \sum_{r} \vec{S}_{r} \cdot \vec{S}_{r+2}.
\label{eq:j1j2}
\end{equation}
The ground state $\vert G(\alpha) \rangle$ is a singlet ($S = 0$) for any $\alpha$. The $\alpha = 0$ limit is the gapless HAF 
with a nondegenerate ground state; Faddeev and Takhtajan used the Bethe ansatz to obtain the exact spectrum of two-spinon triplets and singlets~\cite{faddeev81}. 
The degenerate ground states at $\alpha = 1/2$, the Majumdar-Ghosh (MG) point~\cite{ckm69b}, are the Kekul\'e valence bond diagrams 
$\vert K1 \rangle$ or $\vert K2 \rangle$ in which all spins $S_{2r}$ are singlet paired with either spin $S_{2r+1}$ or $S_{2r-1}$. 
The initial studies~\cite{haldane82, *haldane82_2nd,kuboki87,affleck88,nomura1992} of $H(\alpha)$ focused on the critical point $\alpha_c = 0.2411$ 
at which a spin gap $\Delta(\alpha)$ opens, spin correlations have finite range, and the ground state is doubly degenerate. 
The critical point obtained by level crossing~\cite{nomura1992} has been discussed in terms of field theory and a Kosterlitz-Thouless transition.

The $J_1-J_2$ model at $\alpha > 1$ describes HAFs on sublattices of odd and even numbered sites. It can be viewed~\cite{allen1997,fabian98,affleck96,itoi2001} 
as a zig-zag chain or a two-leg ladder with skewed rungs $J_1$ and rails $J_2$. Now $H(\alpha)/\alpha$ has $J_2 = 1$ and $J_1 = 1/\alpha$ 
is a frustrated interaction between sublattices. The $1/\alpha = 0$ limit of noninteracting HAFs is gapless; the ground state of the 
decoupled phase is nondegenerate with quasi-long-range spin correlations within sublattices. The spin gap $\Delta(\alpha)$ opens at the critical point 
$1/\alpha_2 = 0.44$ and the ground state becomes doubly degenerate~\cite{soos-jpcm-2016}. This critical point is mildly controversial because field 
theories~\cite{allen1997,fabian98,affleck96,itoi2001} with different approximations limit the gapless phase to the point $J_1 = 0$; however, level crossing 
at $\alpha > 2$ was not recognized. The difference between the $\alpha < 1$ and $\alpha^{-1} < 1$ sectors was a motivation for the present study.

Thermal and magnetic fluctuations are suppressed at $T = 0$. The spin gap $\Delta(\alpha)$ is insufficient to characterize how the entropy $S(T,\alpha)$ 
or magnetic susceptibility $\chi(T,\alpha)$ of gapped correlated 1-D systems decreases on cooling. We find power laws $T^{-x(\alpha)}$ that modify 
$\exp(-\Delta(\alpha)/T)$ at low $T < 0.05$. The exponent $x(\alpha)$ depends on frustration: It increases with $\alpha$ for thermal fluctuations 
and decreases with $\alpha$ for magnetic fluctuations. Thermodynamics at $T < 0.05$ is a prerequisite for such results that, as far as we know, 
have not been reported for the $J_1-J_2$ model. Indeed, the low $T$ entropy turns out to be a good way to characterize the model. 

We obtain the thermodynamics by exact diagonalization (ED) of Eq.~\ref{eq:j1j2} in small systems of $N = 4n$ spins and periodic boundary 
conditions followed by density matrix renormalization group (DMRG) calculations of the low-energy states of larger systems of $N \sim 100$ or more~\cite{sudip19}. 
DMRG is powerful numerical method~\cite{white-prl92, *white-prb93}, now well established~\cite{schollwock2005,karen2006}, for the ground state and elementary 
excitations of 1-D models. Convergence to the thermodynamic limit is directly 
seen at $T > T(\alpha,N)$ as thermal fluctuations suppress correlations between distant spins. The full spectrum of $2^N$ spin states is required 
for small systems but not for large ones. Extrapolation to lower $T < T(\alpha,N)$ is possible and makes the thermodynamics accessible 
to $T \sim 0.01 J_1$ for $\alpha < 1$ or to $\sim 0.01 J_2$ for $\alpha^{-1} < 1$.

\begin{figure}[]
         \begin{center} \includegraphics[width=\columnwidth]{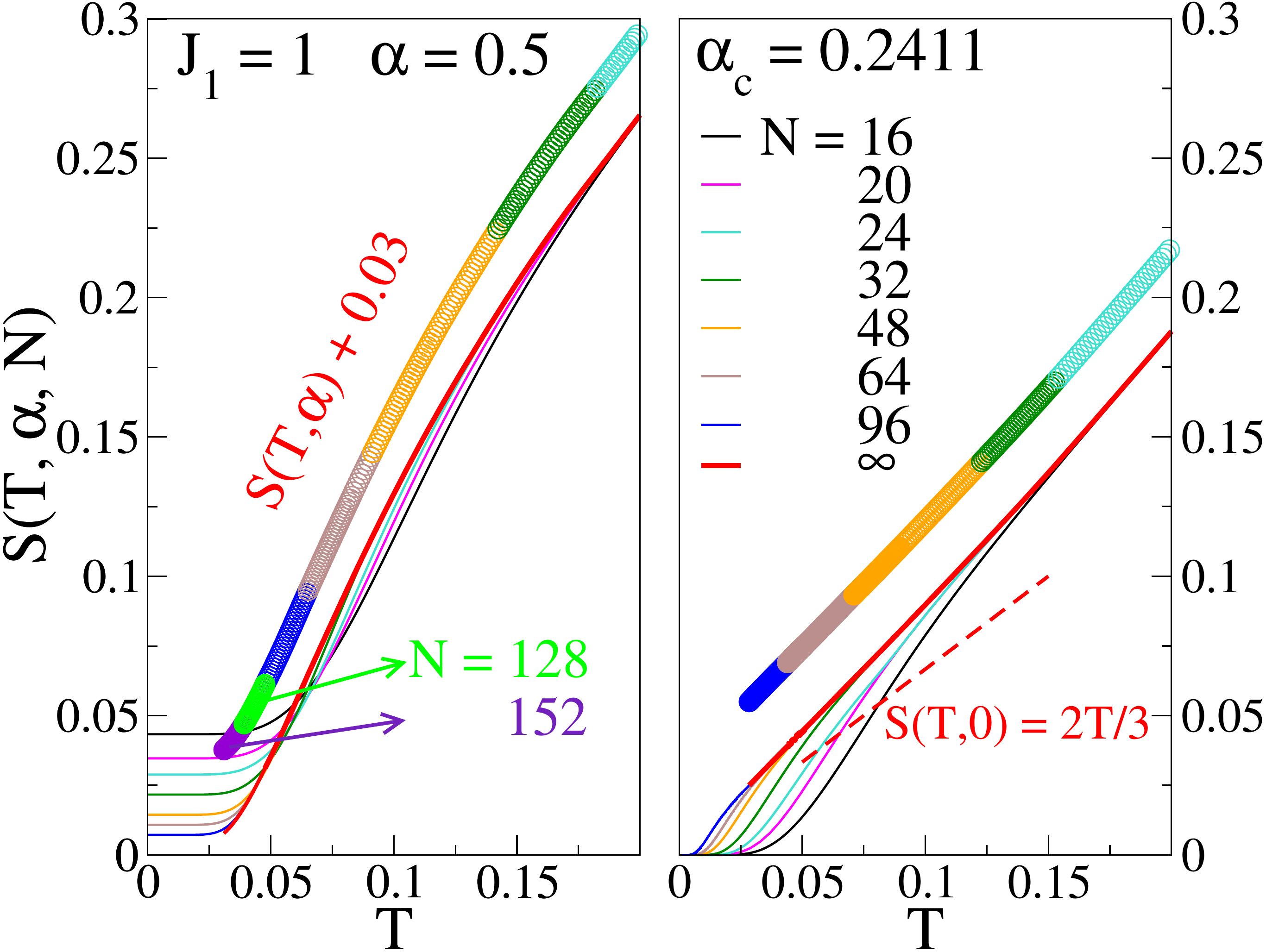}
                 \caption{
Entropy per site $S(T,\alpha,N)$ at $\alpha = 0.50$ (left panel, MG point) and $\alpha_c = 0.2411$ (right panel, critical point) at system size $N$ 
in Eq.~\ref{eq:j1j2}. The thermodynamic limit $S(T,\alpha)$ is the red line that holds for $T > T(\alpha,N)$. $S(T,\alpha)$ is shifted up by 0.03 and 
color coded according to the contributing $N$. The degenerate MG ground state gives $S(0,1/2,N) = N^{-1}\ln2$. Finite size gaps decrease $S(T,\alpha_c,N)$ 
at low $T$. The HAF entropy $S(T,0) = 2T/3$ is exact~\cite{johnstonprl} as $T \rightarrow 0$.                   }
 \label{fig1}
 \end{center}
 \end{figure}

The entropy density $S(T,\alpha)$ illustrates convergence to the thermodynamic limit and differences between gapped and gapless 
quantum phases. The left panel of Fig.~\ref{fig1} shows the entropy per site $S(T,\alpha,N)$ at the MG point where the ground state of 
finite chains is doubly degenerate and $\Delta(1/2)$ is substantial. ED for $N = 16$, 20 and 24 converges from below to $S(T,0.5)$ for $T > 0.15$. 
DMRG for the low-energy states of larger systems extends the limit to $T(1/2,152) \sim 0.03$ as shown by the continuous red line and summarized 
in Section~\ref{sec2}. The converged line is shifted up by $S = 0.03$ and color coded according to the contributing system size; $T(\alpha,N)$ 
is the low-$T$ edge. The ground state degeneracy leads to exactly $N^{-1}\ln2$ at $T = 0$. The thermodynamic limit between $T = 0$ and $T(1/2,152)$ 
is approximated in Section~\ref{sec4}.

The right panel shows the corresponding results for $S(T,\alpha,N)$ at the critical point~\cite{nomura1992} $\alpha_c = 0.2411$ where the gap 
$\Delta(\alpha)$ opens. The ground state of finite systems is nondegenerate except at $\alpha = 1/2$. Calculations to $N = 96$ return the thermodynamic 
limit for $T(\alpha_c,96) > 0.025$, below which finite size gaps are evident. The color-coded line $S(T,\alpha_c)$ is again shifted by 0.03. 
The dashed line $S(T,0)$ is the exact~\cite{johnstonprl} HAF limit, initially linear in $T$, that previously served to validate the ED/DMRG 
method~\cite{sudip19}. Frustration increases $S(T,\alpha_c)$ by about $20\%$ above $S(T,0)$ at low $T$. Extrapolation yields the 
thermodynamic limit for $T < T(\alpha_c,96)$.

Since $S(T,\alpha)$ is linear in gapless 1D chains, $S^\prime(0,\alpha)$ is finite up to $\alpha_c$ while $\Delta(\alpha) > 0$ 
ensures $S^\prime(0,\alpha) = 0$ in gapped chains. Entropy calculations provide an independent new way of estimating quantum critical 
points. Frustration increases the density of states at low $T$ compared to the HAF while 
$\Delta(1/2)$ initially decreases $S(T,1/2)$ 
at the MG point.

Two-spin correlation functions at frustration $\alpha$ are ground state expectation values,
\begin{equation}
	C_2(p,\alpha)=\langle G(\alpha) \vert \vec{S}_1 \cdot \vec{S}_{p+1} \vert G(\alpha) \rangle \equiv 3 \langle  S^z_1 S^z_{p+1} (\alpha) \rangle.
\label{eq:2spincor}
\end{equation}
We have used periodic boundary conditions and isotropic exchange in Eq.~\ref{eq:2spincor}. HAF correlations $C_2(p,0)$ 
are exact~\cite{sato2005} up to $p = 5$; they are quasi-long-ranged and go~\cite{affleck89,sandvik2010} as $(-1)^p(\ln p)^{1/2} /p$ for $p \gg 1$. 
$C_2(p,\alpha)$ is quasi-long-ranged up to $\alpha_c$. The range then decreases to first neighbors at the MG point where $C_2(p,1/2) = 0$ for $p \ge 2$. 
The $\alpha^{-1} = 0$ limit of HAFs on sublattices has vanishing correlations for odd $p$ for spins in different sublattices and quasi-long-range 
correlations for even $p$. The $\alpha < 1$ and $\alpha^{-1} < 1$ sectors have different but related spin correlations.

The paper is organized as follows. The ED/DMRG method is summarized in Section~\ref{sec2} using the size dependence of the magnetic 
susceptibility $\chi(T,\alpha,N)$ and entropy $S(T,\alpha,N)$ per site. The energy spectrum $\{E(\alpha,N)\}$ of Eq.~\ref{eq:j1j2} and 
partition function $Q(T,\alpha,N)$ yield the thermodynamics. The entropy and spin specific heat $C(T,\alpha)$ are obtained in Section~\ref{sec3} in 
gapless chains and approximated in gapped chains. We find  the inflection point $T^*(\alpha)$ of $S(T,\alpha)$ and relate it to the power 
law $T^{-\gamma(\alpha)}$ that modifies $\exp(-\Delta(\alpha)/T)$. Section~\ref{sec4} presents the thermodynamic determination of critical 
points and differences between intrachain frustration leading to $\alpha_c$ and interchain frustration leading to $1/\alpha_2$. 
Converged susceptibilities $\chi(T,\alpha)$ are reported in Section~\ref{sec5} for gapless and gapped chains. They are modeled 
using $T^*(\alpha)$ and the power law $T^{-\eta(\alpha)}$. The ratio $R(T,\alpha) = S(T,\alpha)/4T\chi(T,\alpha)$ is the relative 
contribution of thermal and magnetic fluctuations. It is initially constant in gapless chains and focuses attention on deviations 
from $\exp(-\Delta(\alpha)/T)$ in gapped chains. $R(T,\alpha)$ decreases on cooling below $0.05 T$ for $\alpha = 0.35$ or 0.40, and 
it increases for $\alpha = 0.50$ or 0.67. Section~\ref{sec6} is brief discussion and summary.

\section{\label{sec2} DMRG and convergence}
The molar magnetic susceptibility $\chi(T,\alpha)$ provides direct comparison with experiment since electronic spins dominate 
the magnetism. The reduced susceptibility is in units of $N_Ag^2 \mu_B^2/J_1$ where $N_A$ is the Avogadro constant, $\mu_B$ is the 
Bohr magneton and $g = 2.00232$ is the free-electron $g$ factor. Isotropic exchange rules out spin-orbit coupling. We take $J_1 = 1$ or $J_2 = 1$, 
respectively, for $\alpha < 1$ or $\alpha^{-1} < 1$ calculations. The energy spectrum $\{E(\alpha,N)\}$ of $H(\alpha)$ has $2^N$ spin 
states. Given $\{E(\alpha,N)\}$, the partition function $Q(T,\alpha,N)$ is the sum over $\exp(-\beta E_p(\alpha,N))$, with $\beta = 1/k_BT$ 
and Boltzmann constant $k_B$. Standard statistical mechanics yields $\chi(T,\alpha,N)$, $S(T,\alpha,N)$, $C(T,\alpha,N)$, and 
spin correlation functions $C_2(T,p,\alpha)$ of finite systems.

\begin{figure}[]
         \begin{center} \includegraphics[width=\columnwidth]{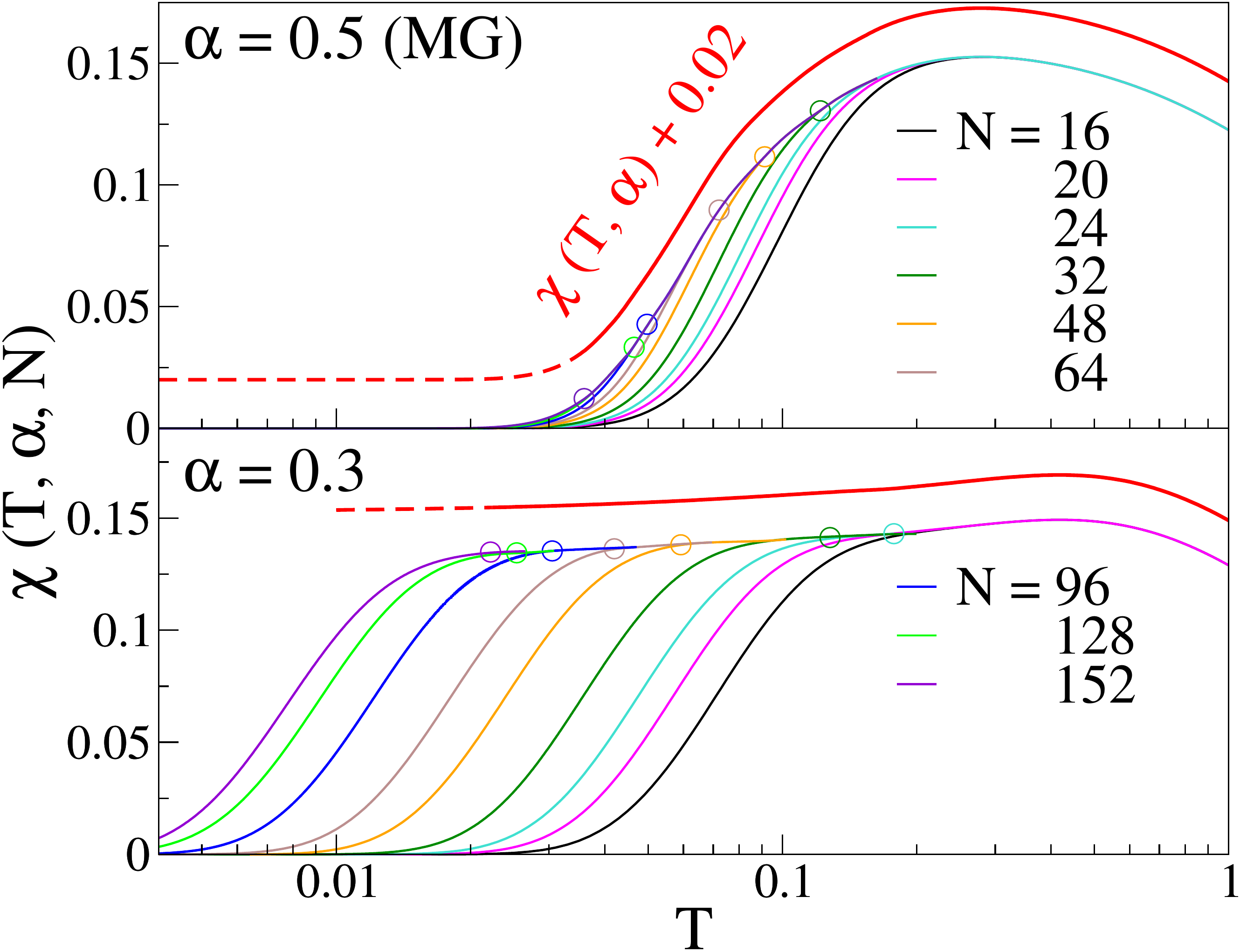}
                 \caption{
Molar magnetic susceptibility $\chi(T,\alpha,N)$ of Eq.~\ref{eq:j1j2} with $N$ spins at $\alpha = 0.50$ and 0.30. Convergence to the $\chi(T,\alpha)$ 
line, shifted up by $0.02$ in both panels, holds for $T > T(\alpha,N)$, the 
open circles. The continuous and dashed red lines are converged $\chi(T,\alpha)$ and 
extrapolation, respectively, to $T < T(\alpha,152)$.  
		 }
 \label{fig2}
 \end{center}
 \end{figure}

We discuss the ED/DMRG method by following the convergence of $\chi(T,\alpha,N)$ in Fig.~\ref{fig2} to $\chi(T,\alpha)$ with increasing system 
size at $\alpha = 0.50$ and $0.30$. The logarithmic scale focuses attention on low $T$. The solid red lines are $\chi(T,\alpha) + 0.02$, displaced 
upwards from the finite-size calculations. ED of Eq.~\ref{eq:j1j2} up to $N = 24$ demonstrates convergence for $T > 0.2$ in either case using the 
full spectrum of $2^N$ states. DMRG returns the low-energy states $E_p(\alpha,N)$ of larger systems. Finite gaps $\Delta(\alpha,N)$ to the lowest 
triplet decrease with $N$ and suppress the susceptibility at $T = 0$. Convergence to the thermodynamic limit requires $N^{-1} \ll \Delta(\alpha)$, 
a condition that is almost satisfied at $N = 96$, $128$ or $152$ in the upper panel. The exponentially small gap at $\alpha = 0.30 > \alpha_c$ 
is not at all evident in the lower panel even at $N = 152$. 

We summarize the DMRG calculations in sectors with total $0 \le S^Z \le N/2$ presented in detail and tested in Ref.~\onlinecite{sudip19}. 
The singlet ground state is in the $S^Z = 0$ sector. We use periodic boundary conditions, increase the system size by four spins at each step of 
infinite DMRG, and keep $m = 500$ eigenstates of the system block. The total dimension of the superblock (the Hamiltonian matrix) is 
approximately $500 \times 500 \times 4 \times 4$ ($\sim 10^6$). Varying $m$ between 300 and 500 indicates a $3-4$ decimal place accuracy 
of low-lying levels, which are explicitly known for the HAF ($\alpha = 0$) at system size $N$. We target the lowest few hundred of states in $S^Z$ 
sectors instead of the ground state and energy gaps in standard DMRG.

We introduce a cutoff with $E_p(\alpha,N) \le E_C(\alpha,N)$ and compute the entropy per site $S_C(T,\alpha,N)$ of the truncated spectrum. 
Increasing $E_C(\alpha,N)$ ensures convergence to $S(T,\alpha,N)$ from below since truncation should not reduce the entropy. We increase 
the cutoff until the maximum of $S_C(T,\alpha,N)/T$ has converged or almost converged. The maxima $T(\alpha,N)$ are shown as open points 
in Fig.~\ref{fig2}. The entropy at $T = T(\alpha,N)$ is the best approximation to $S(T,\alpha)$ for the cutoff. The truncated spectrum 
suffices for a small interval $T \ge T (\alpha,N)$ of converged thermodynamics at each system size before truncation takes its toll; 
additional points $T(\alpha,N)$ can be found. The thermodynamic limit $\chi(T,\alpha) + 0.02$ is shown as a bold red line through the 
points that smoothly connects to ED at high $T$. Convergence to $\chi(T,\alpha)$ is from below and has been checked~\cite{sudip19} 
against the exact HAF susceptibility.

%We summarize the ED/DMRG method by following the convergence of $\chi(T,\alpha,N)$ in Fig.~\ref{fig2} to $\chi(T,\alpha)$ with 
%increasing system size at $\alpha = 0.50$ and 0.30. The logarithmic scale focuses attention on low $T$. ED of Eq.~\ref{eq:j1j2} up to $N = 24$ 
%demonstrates convergence for $T > 0.2$ in either case using the full spectrum of $2^N$ states. DMRG returns the low-energy states $E_p(\alpha,N)$ 
%of larger systems. Finite gaps $\Delta(\alpha,N)$ to the lowest triplet decrease with $N$ and suppress the susceptibility at $T = 0$. 
%Convergence to the thermodynamic limit requires $N^{-1} \ll \Delta(\alpha)$, a condition that is almost satisfied at $N = 96$, $128$ or $152$ in the 
%upper panel. The exponentially small gap at $\alpha = 0.30 > \alpha_c$ is not at all evident in the lower panel even at $N = 152$.

%
%

%DMRG calculations in the context of thermodynamics target the lowest hundreds of states $E_p(\alpha,N)$ instead of the ground state or $T = 0$ 
%energy gaps. The computational details are given Ref.~\onlinecite{sudip19}. We use periodic boundary conditions and increase the system size by four 
%spins at each step of infinite DMRG. We introduce a cutoff with $E_p(\alpha,N) \le E_C(\alpha,N)$ and obtain the entropy per site $S_C(T,\alpha,N)$ 
%of the truncated spectrum. Increasing $E_C(\alpha,N)$ ensures convergence to $S(T,\alpha,N)$ from below since truncation cannot reduce the entropy. 
%We increase the cutoff until the maximum of $S_C(T,\alpha,N)/T$ has converged or almost converged. The maxima $T(\alpha,N)$ are shown as open points 
%in Fig.~\ref{fig2}. The entropy at $T = T(\alpha,N)$ is the best approximation to $S(T,\alpha)$ for the cutoff. The truncated spectrum 
%suffices for a small interval $T \ge T(\alpha,N)$ of converged thermodynamics at each system size. 
%The $\chi(T,\alpha,N)$ lines are shown to $T = T(\alpha,N')$ where $N'$ is next smaller system. The thermodynamic limit $\chi(T,\alpha)$ is 
%shown as a bold red line through the points, shifted up by $0.02$ that smoothly connects to ED at high $T$. Convergence to $\chi(T,\alpha)$ is from below and has been 
%checked~\cite{sudip19} against the exact HAF susceptibility.

The DMRG results for $S(T,\alpha,N)$ in Fig.~\ref{fig1} for $N > 24$ are also based on truncated $E_p(\alpha,N) \le E_C(\alpha,N)$. They converge for $T > T(\alpha,N)$ 
at the lower edges of the colored-coded line. The procedure is general. Other systems sizes, including larger ones, can be studied. The numerical accuracy 
is ultimately limited by the density of low-energy states of large systems~\cite{sudip19}. ED/DMRG exploits the fact that a few hundreds of 
states $E_p(\alpha,N) \le E_C(\alpha,N)$  
in sectors with $S^Z = 0, 1, ...$ suffice
for the thermodynamics in a limited range of $T$ at each system size. The discarded 
states have Boltzmann factors with $\beta E_C(\alpha,N) > 10$ in the following results.

\begin{figure}[]
         \begin{center} \includegraphics[width=\columnwidth]{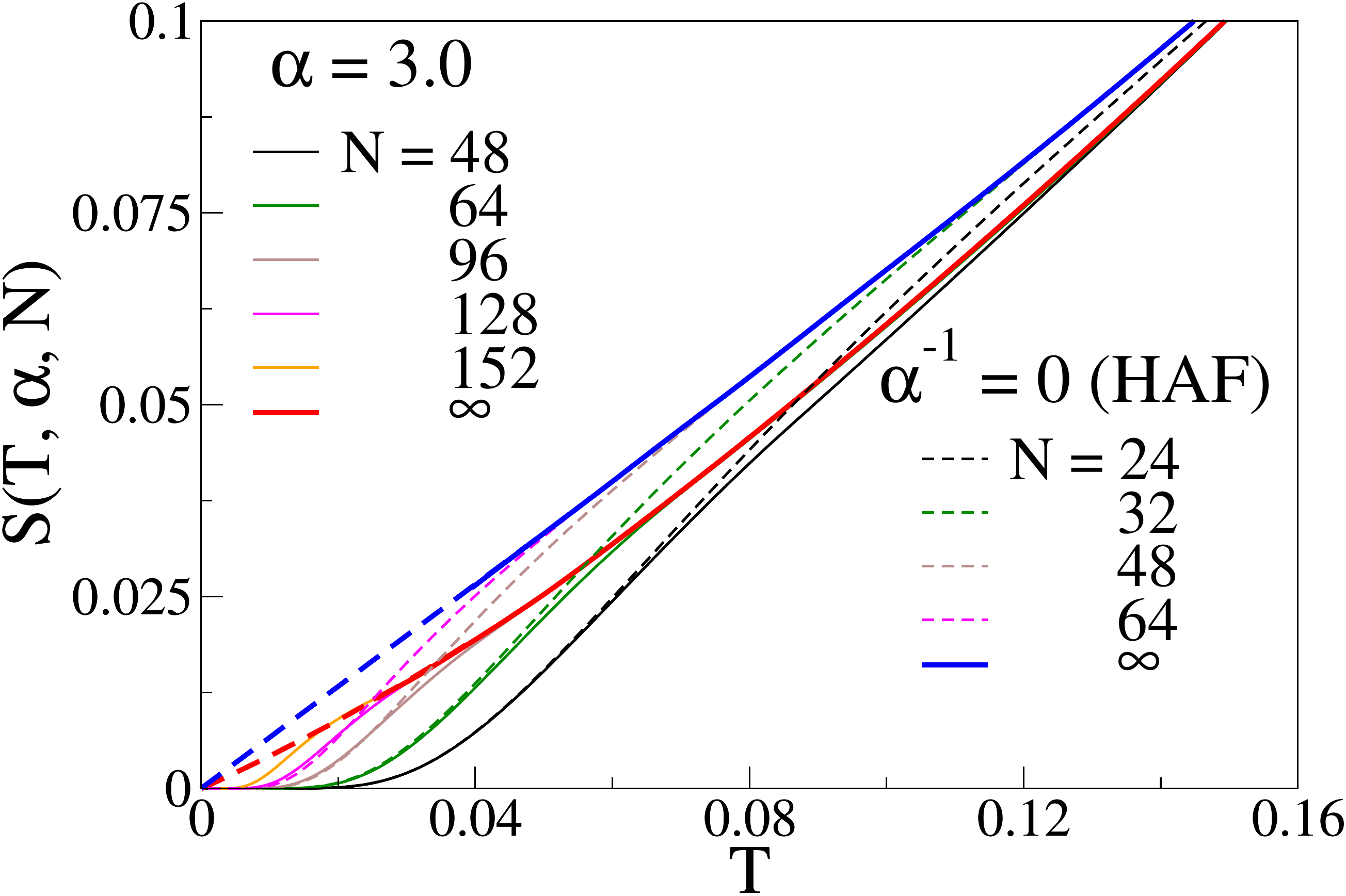}
                 \caption{
			 Entropy density $S(T,\alpha,N)$ of Eq.~\ref{eq:j1j2} with $J_2 = 1$, $J_1 = 1/3$ and system size $N$ (continuous lines) 
			 and with $J_2 = 1$, $J_1 = \alpha^{-1} = 0$ and $N/2$ (dashed lines). Converged $S(T,\alpha)$ are shown as continuous 
			 bold lines, and extrapolation to $T = 0$ as dashed bold lines.}
 \label{fig3}
 \end{center}
 \end{figure}

Convergence to the thermodynamic limit is more challenging in the $\alpha^{-1} < 1$ sector of weak exchange between HAFs with $J_2 = 1$ in sublattices. 
The system size is effectively $N/2$ instead of $N$. We compare in Fig.~\ref{fig3} the entropy densities $S(T,\alpha,N)$ at $\alpha^{-1} = 1/3$ 
and system size $N$ with $\alpha^{-1} = 0$ (HAF) and $N/2$. Interchain exchange $\alpha^{-1} = 1/3$ hardly changes the entropy of finite systems 
below $T = 0.06$. Moreover, interchain exchange reduces the entropy compared to $S(T,0)$ while $\alpha_c$ in Fig.~\ref{fig1}, right panel, increases 
the entropy. These qualitative differences are related to spin correlation functions. We obtain convergence to the thermodynamic limit 
for $T > T(\alpha,N) = 0.023$ for $\alpha = 3$, $N = 152$. 

The initial ED/DMRG calculations were up to system size $N = 96$ and returned converged thermodynamic for $T > T(\alpha,96)$. 
About half of the calculations were subsequently extended to $N = 128$ or $152$ and convergence for $T > T(\alpha,152)$ in order to address specific points. 
Converged results are shown as continuous lines down to $T(\alpha,96)$ and $T(\alpha,152)$ for $S(T,\alpha)$ in Fig.~\ref{fig1} and $\chi(T,\alpha)$ in Fig.~\ref{fig2}, respectively. 

It has been very instructive to follow the size dependence of thermodynamic quantities explicitly to suggest possible extrapolation or interpolation to lower $T$. 
Larger $N \sim 200$ is accessible with sufficient motivation. We know on general grounds that $S(0,\alpha) = 0$ and that gapped systems have $S^\prime(0,\alpha) = 0$. 
The thermodynamic limit of the entropy in Figs. ~\ref{fig1} or ~\ref{fig3} is obtained more accurately than the magnetic susceptibility in Fig.~\ref{fig2}. 
It turns out that $S(T,\alpha)$ is an effective way to characterize the low $T$ thermodynamics of the $J_1-J_2$ model, Eq.~\ref{eq:j1j2}.

\section{\label{sec3} Entropy and specific heat}

\begin{figure}[]
         \begin{center} \includegraphics[width=\columnwidth]{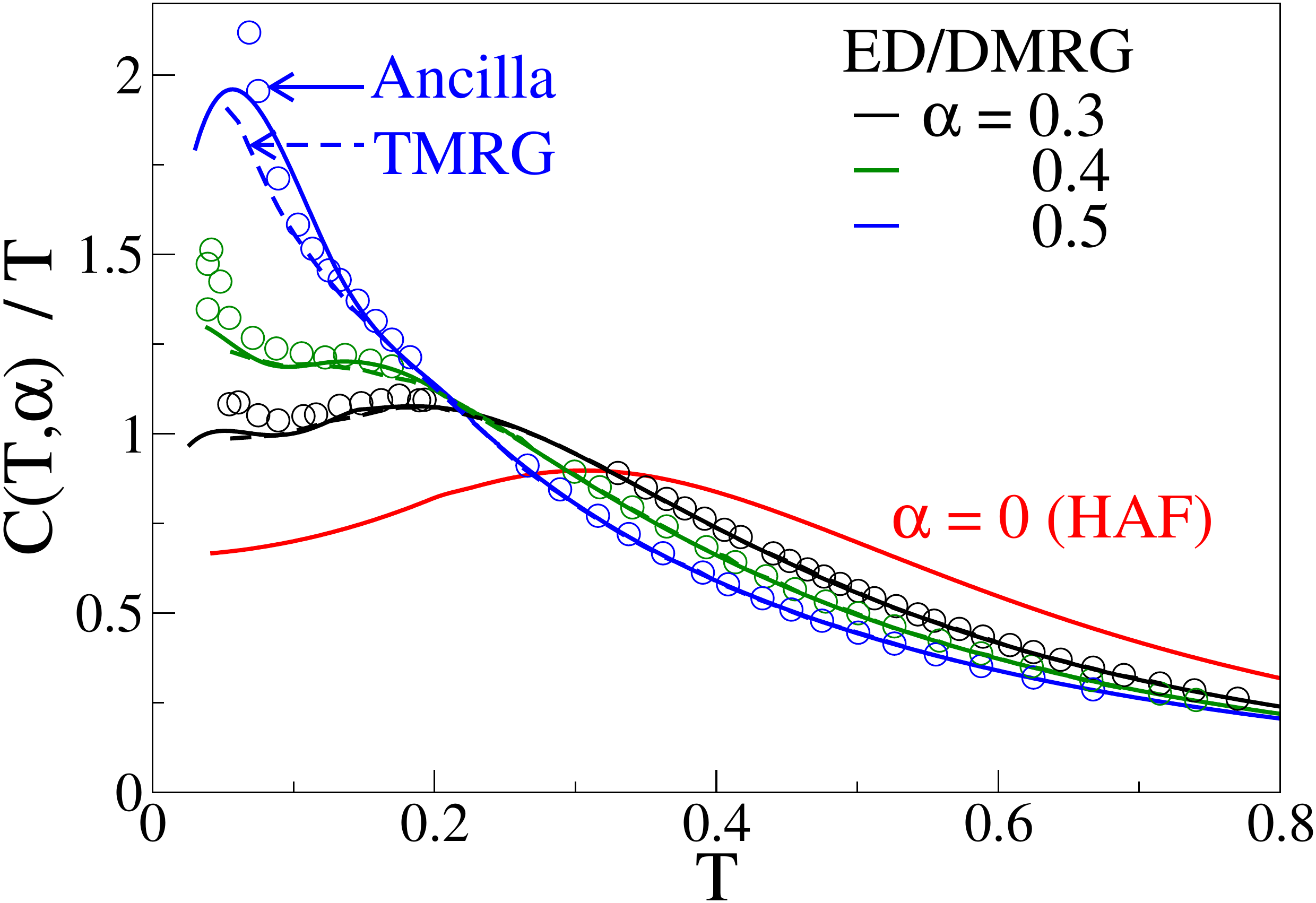}
                 \caption{Thermodynamic limit of the entropy derivative $S^\prime(T,\alpha) = C(T,\alpha)/T$ from $T = 0.05$ 
		 to 0.8 at frustration $\alpha$ in Eq.~\ref{eq:j1j2}. Continuous lines are ED/DMRG; dashed lines are 
		 TMRG, Fig. 5(b) of Ref.~\onlinecite{okunishiprb}; points are ancilla calculations, Fig. 3(a) of Ref.~\onlinecite{white2005}. 
		 Differences appear at $T < 0.10$.}
 \label{fig4}
 \end{center}
 \end{figure}

We obtain in this Section the entropy density $S(T,\alpha)$ of the $J_1-J_2$ model, Eq.~\ref{eq:j1j2}, at low $T$. Converged $S(T,\alpha)$ 
gives the spin specific heat $C(\alpha,T)$ per site as the derivative $S^\prime(T,\alpha) = C(T,\alpha)/T$. ED to system size $N = 24$ 
and DMRG to $N = 96$ return converged $S(T,\alpha)$ for $T > 0.15$ and $T > T(\alpha,96)$, respectively. The continuous lines in 
Fig.~\ref{fig4} are calculated $C(T,\alpha)/T$ at the indicated $\alpha$ and $T > T(\alpha,96)$. Frustration increases the $S^\prime(T,0)$ 
maximum of the HAF and shifts it to lower $T$. The TMRG results in Fig. 5(b) of Ref.~\onlinecite{okunishiprb} extend down to $T/J_1 = 0.05$. 
DMRG results with an expanded Hilbert space and ancilla are shown down to $T/J_1 = 0.05$ in Fig. 3(a) of Ref.~\onlinecite{white2005}. 
The $C(T,\alpha)/T$ curves agree quantitatively for $T > 0.2$ where the thermodynamic limit is now accessible by ED. There are differences 
at low $T$. For example, the previous $C(T,0.5)/T$ curves increase continuously down to $T = 0.05$ while we find a maximum. A maximum appears~\cite{okunishiprb} 
at $\alpha = 0.6$ with larger spin gap. We seek the thermodynamics below $T \sim 0.1$.

\begin{figure}[]
         \begin{center} \includegraphics[width=\columnwidth]{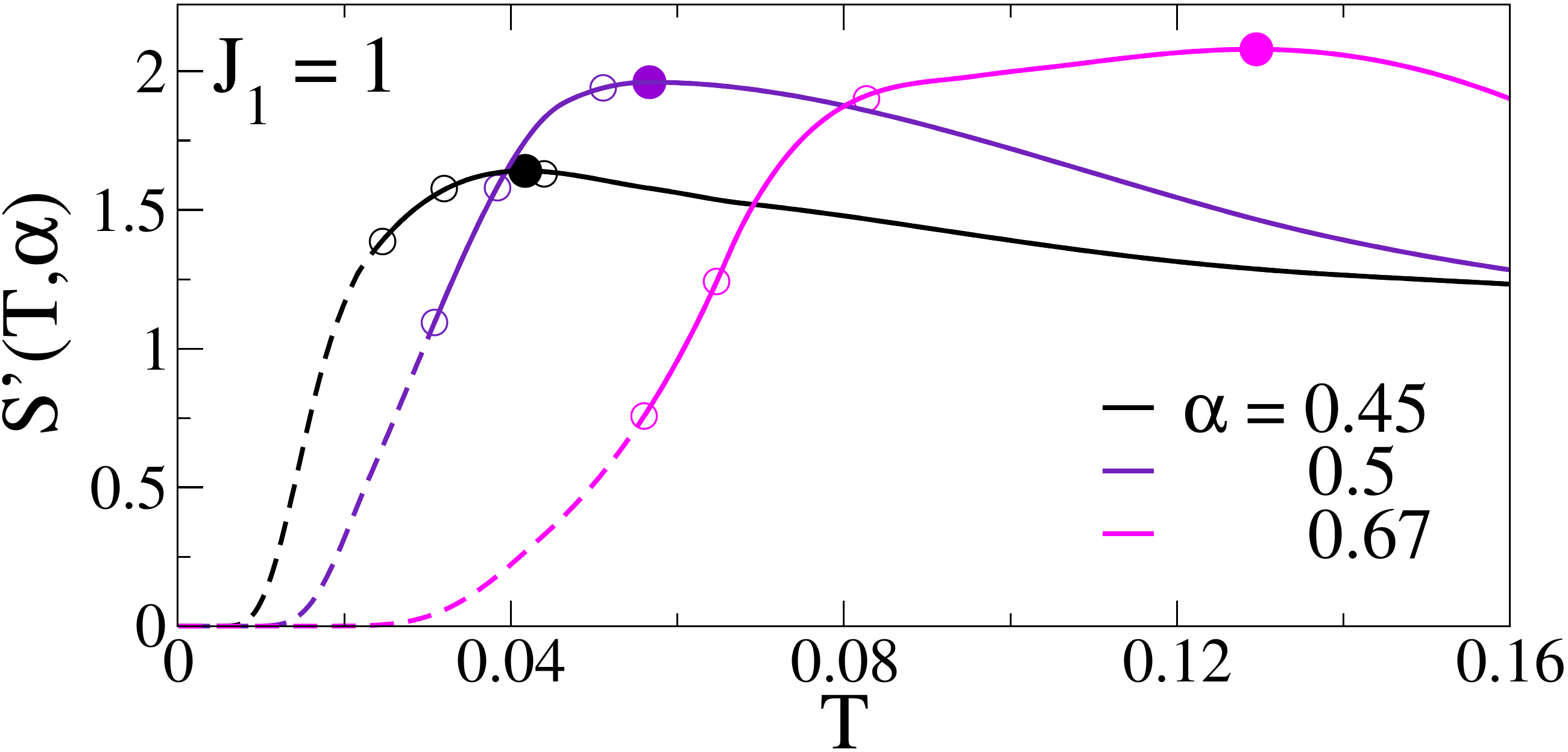}
                 \caption{$S^\prime(T,\alpha)$ to $T = 0.16$ for $\alpha = 0.45$, 50 and 0.67. Continuous lines are converged results for $T > T(\alpha,152)$. 
		 Open points are $T(\alpha,N)$ at $N = 96$, 128 and 152; solid points are the maxima $T^*(\alpha)$. The dashed lines are Eq.~\ref{eq:apprxen} up to $T(\alpha,152)$.}			 
 \label{fig5}
 \end{center}
 \end{figure}

Turning to low $T$, we show $S^\prime(T,\alpha)$ results in Fig.~\ref{fig5} for systems with large spin gaps $\Delta(\alpha)$. 
Open points at $T(\alpha,N)$ mark converged $S^\prime(T,\alpha)$ for $\alpha = 0.45$, 0.50 and 0.67 at system size $N = 96$, 
128 and 152. The $S^\prime(T,\alpha)$ maxima at $T^*(\alpha)$ are directly accessible when $N^{-1} \ll \Delta(\alpha)$. They 
are points of inflection where the curvature $S^{\prime\prime}(T^*,\alpha)$ is zero. Since gapped chains have $S^\prime(0,\alpha) = 0$, 
they necessarily have $T^*(\alpha) > 0$. However, exponentially large $N$ will be needed to resolve $T^*(\alpha)$ when the gap is 
exponentially small. The dashed lines in Fig.~\ref{fig5} are based on a phenomenological approximation. We discuss the entropy of gapless chains and gapped chains with $T^*(\alpha) < T(\alpha,N)$ for the largest system studied before returning to the dashed lines in Fig.~\ref{fig5}.

\begin{figure}[h]
         \begin{center} \includegraphics[width=\columnwidth]{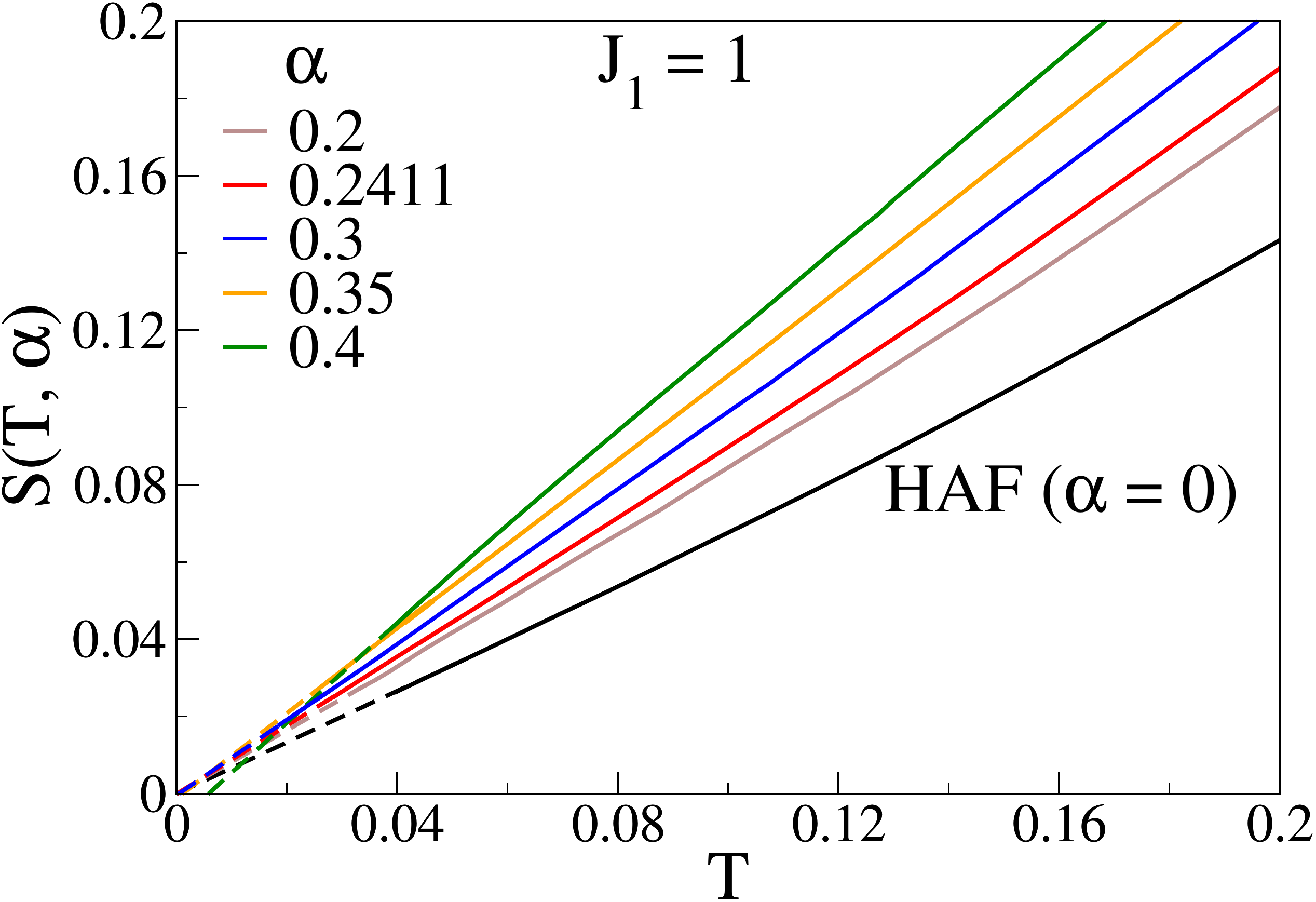}
                 \caption{
		Entropy density $S(T,\alpha)$ up to $T = 0.2$ at frustration $\alpha$ in Eq.~\ref{eq:j1j2}. Continuous lines for $T > T(\alpha,N)$, 
		 $N = 96$  or $N = 152$ for $\alpha = 0.30$. Dashed lines are linear extrapolation using Eq.~\ref{eq:linearen}. 
		 }
 \label{fig6}
 \end{center}
 \end{figure}

The entropy is strikingly different in chains with small or no spin gap. Fig.~\ref{fig6} shows converged $S(T,\alpha)$ up to $T = 0.20$ 
and frustration $\alpha$. Continuous lines are DMRG results for $T \ge T(\alpha,N)$ with $N = 96$, except for $N = 152$ at $\alpha = 0.30$. 
They are model exact and initially linear in $T$ in gapless chains with $S^\prime(0,\alpha) > 0$. Small $\Delta(\alpha)$ at $\alpha > \alpha_c$ 
enforces $S^\prime(0,\alpha) = 0$  without otherwise spoiling the linear regime. The dashed lines $T \le T(\alpha,N)$ are linear extrapolations
\begin{equation}
	S(T,\alpha)=A(\alpha)T-B(\alpha)
	\label{eq:linearen}
\end{equation}
based on the calculated $A(\alpha)$. The linear regime has $S^{\prime\prime}(T,\alpha) = 0$ over an interval that shrinks to a point of 
inflection $T^*(\alpha)$ with increasing $\Delta(\alpha)$ at the $S^\prime(T,\alpha)$ maxima in Fig.~\ref{fig5}. It follows that Eq.~\ref{eq:linearen} is limited to
some $\alpha < 0.45$ that remains open.

The linear regime with $B(\alpha) = 0$ extends to $T = 0$ in gapless chains. In chains with a small gap, $S(T,\alpha)$ 
is initially linear at $T > T(\alpha,N)$, here $N = 96$, and presumably to $T^*(\alpha) < T(\alpha,96)$ in longer chains. 
The functional form at low $T$ is not known. As a simple phenomenological approximation, we take
\begin{equation}
	S(T,\alpha)=c(\alpha)T^{-\gamma(\alpha)}\exp(-\Delta(\alpha)/T).
        \label{eq:apprxen}
\end{equation}
The range is from $T = 0$ to $T^*(\alpha)$ or $T(\alpha,N)$, whichever is lower, where $T(\alpha,N)$ refers to the largest system 
studied. We match the magnitude and slope at $T(\alpha,N)$ when $T(\alpha,N) < T^*(\alpha)$ to find $\gamma(\alpha)$ and $c(\alpha)$. 
When $T^*(\alpha) < T(\alpha,N)$, we extrapolate Eq.~\ref{eq:linearen} to lower $T$ and find $c(\alpha)$, $\gamma(\alpha)$ and $T^*(\alpha)$ 
by setting $S^{\prime\prime}(T^*,\alpha) = 0$ and matching the magnitude and slope of the extrapolated $S(T,\alpha)$ at $T^*(\alpha)$.

\begin{table}[b]
\caption{\label{tab1}
Singlet-triplet gap $\Delta(\alpha)$, entropy parameters $A(\alpha)$ and $B(\alpha)$ in Eq.~\ref{eq:linearen} at frustration $\alpha$
        in Eq.~\ref{eq:j1j2} and $T(\alpha,96)$ discussed in the text.}
\begin{ruledtabular}
\begin{tabular}{ c  c  c  c  c }
$\alpha$ &  $ \Delta(\alpha)$ &  $A(\alpha)$  &   $B(\alpha)$  &  $T(\alpha,96)$      \\ \hline

	0.4    &	0.0299 &	1.292 &	0.0075 &	0.039 \\
	0.35 &	0.0053 &	1.102 &	0.0012 &	0.025  \\
	0.3 &	0.00074	 & 0.980	 & 0.00056 &	0.025  \\
	0.2411$^{\rm a}$  &	0 &	0.885 &	0.00008 &	0.029 \\
	0.2 &	0 &	0.820 &	0 &	0.033 \\
 0$^{\rm b}$ &	0 &	0.663 &	0 &	0.039 \\

\end{tabular}
\end{ruledtabular}
$^{\rm a}$ critical point; $^{\rm b}$ HAF \qquad \qquad  \qquad \qquad \qquad \qquad \qquad \qquad \qquad
\end{table}

The $\alpha \ge 0.45$ systems in Fig.~\ref{fig5} have converged $S(T,\alpha)$ for $T \ge T(\alpha,152)$ and resolved $S^\prime(T,\alpha)$ 
maximum $T^*(\alpha)$.  Matching slopes at $T = T(\alpha,N)$ leads to
\begin{equation}
	\gamma(\alpha)=\frac{\Delta(\alpha)}{T}-\frac{TS^\prime(T,\alpha)}{S(T,\alpha)}.
        \label{eq:gamma}
\end{equation}
At $T(\alpha,152)$ we find $TS^\prime/S = 2.63$ at $\alpha = 0.45$, 4.47 at 0.50 and 3.80 at $\alpha = 0.67$. 
Spin gaps $\Delta(\alpha)$ are obtained by $1/N$ extrapolation of DMRG gaps $\Delta(\alpha,N)$ in chains up to $N = 100$. 
They are 0.113, 0.233 and 0.433 with increasing $\alpha$. The dashed lines in Fig.~\ref{fig5} up to $T(\alpha,152)$ are 
Eq.~\ref{eq:apprxen} with $\Delta(\alpha)$ and exponent $\gamma(\alpha)$ in Eq. \ref{eq:gamma}. The exponents $\gamma(\alpha)$ 
depend on the system size because Eq.~\ref{eq:apprxen} approximates $S(T,\alpha)$ up to $T(\alpha,N)$. We are interested 
in the dependence of $\gamma(\alpha)$ on frustration rather than its magnitude. Deviations from $\exp(-\Delta(\alpha)/T)$ up 
to, say, $T^*(\alpha)$ clearly require a function with many more parameters than $c(\alpha)$ and $\gamma(\alpha)$.

Linear $S(T,\alpha)$ in Fig.~\ref{fig6} extends to $T^*(\alpha) < T(\alpha,96)$ in systems with $\alpha \le 0.40$. 
The maximum at $S^{\prime\prime}(T^*,\alpha) = 0$ requires large $N$ when $\Delta(\alpha)$ is small. We extrapolate $S(T,\alpha)$ to $T^*(\alpha)$ 
and use Eq.~\ref{eq:apprxen} for $T \le T^*(\alpha)$. Zero curvature at $T^*(\alpha)$ relates the gap and exponent
\begin{equation}
\frac{\Delta(\alpha)}{T^*(\alpha)}=1+\gamma(\alpha)+\sqrt{1+\gamma(\alpha)}
        \label{eq:tstar}
\end{equation}
The coefficient $A(\alpha)$ and $B(\alpha)$ in Eq.~\ref{eq:linearen} are constant in the linear regime. The 
ratio of the slope and the magnitude of $S(T,\alpha)$ at $T^*(\alpha)$ leads to 
\begin{equation}
	\frac{\Delta(\alpha)}{T^*(\alpha)}=y(\alpha)(y(\alpha)-1)
        \label{eq:gaptstar}
\end{equation}
where $y(\alpha)^2 = A(\alpha)\Delta(\alpha)/B(\alpha)$ and $y(\alpha) - 1 = (\gamma(\alpha) + 1)^{1/2}$. 

We discuss $C(T,\alpha)$ at weak frustration $\alpha \le 0.40$ using the coefficients $A(\alpha)$ and $B(\alpha)$ in Eq.~\ref{eq:linearen} and $T(\alpha,96)$. 
Table~\ref{tab1} lists $\Delta(\alpha)$, $A(\alpha)$ and $B(\alpha)$ for both gapless and gapped chains. We find $A(0) = 0.663$ instead of the 
exact~\cite{johnstonprl} $A(0) = 2/3$. The gap opens at $\alpha_c$ and is still tiny at $\alpha = 0.30$. The inferred $T^*(\alpha)$ and $\gamma(\alpha)$ 
based on Eq.~\ref{eq:apprxen} up to $T^*(\alpha)$ are in Table~\ref{tab2}. We have omitted $\alpha = 0.30$, which requires greater 
numerical accuracy, larger $N$, and most likely has $T^*(0.30) < 0.001$. We have included systems with $\alpha \ge 0.45$ and $T^*(\alpha) > T(\alpha,152)$. 
The exponent $\eta(\alpha)$ is obtained later from the susceptibility $\chi(T,\alpha)$.

\begin{table}
\caption{\label{tab2}
$S^\prime(T,\alpha)$ maximum $T^*(\alpha)$ in gapped $J_1-J_2$ models, ratio $\Delta(\alpha)/ T^*(\alpha)$,
        and exponents $\gamma(\alpha)$ in Eq.~\ref{eq:apprxen} and $\eta(\alpha)$ in Eq.~\ref{eq:apprxchi}.
	}
\begin{ruledtabular}
\begin{tabular}{ c  c  c  c  c }
$\alpha$ &  $ T^*(\alpha) $ &  $\Delta(\alpha)/T^*(\alpha) $  &   $\gamma(\alpha)$  &  $\eta(\alpha)$ \\ \hline
        0.67 &  0.130 & 3.34 &  3.56 &  1.24  \\
        0.50$^{\rm a}$ &        0.057 & 4.12 &  4.34 &  1.23 \\
        0.45 &  0.042 & 2.71 &  1.97 &  2.79 \\ 
        0.4 &   0.0104 &        2.88 &  0.61 &  2.71 \\
        0.35 &  0.0020 &        2.65 &  0.46 &  2.61 \\
\end{tabular}
\end{ruledtabular}
$^{\rm a}$ MG point.   \qquad \qquad  \qquad \qquad \qquad \qquad \qquad \qquad \qquad
\end{table}
%

%%%%

\begin{figure}[]
         \begin{center} \includegraphics[width=\columnwidth]{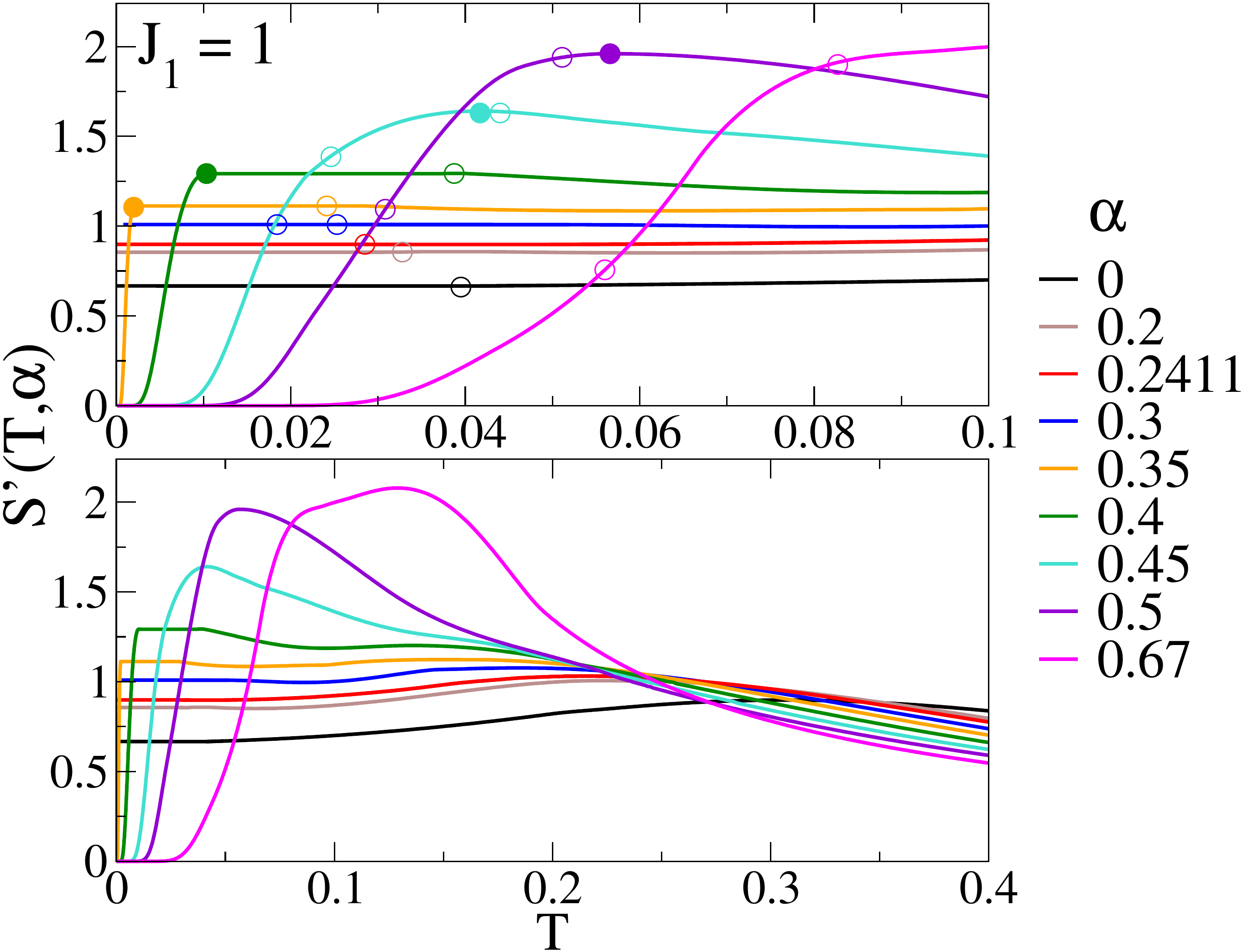}
                 \caption{
$S^\prime(T,\alpha) = C(T,\alpha)/T$ at the indicated $\alpha$ to $T = 0.1$ (upper panel) and $T = 0.4$ (lower panel). 
		 Upper panel: Solid points are $T^*(\alpha)$ in Table~\ref{tab2} for gapped chains with $\alpha \ge 0.35$. 
		 Open points are $T(\alpha,96)$ in Table~\ref{tab1}, $T(\alpha,152)$ at 
$\alpha = 0.30$ and $T(\alpha,N)$ at N = 96, 128 and 152 for $\alpha \ge 0.45$.}

 \label{fig7}
 \end{center}
 \end{figure}

The evolution of $S^\prime(T,\alpha) = C(T,\alpha)/T$ with frustration is shown in Fig.~\ref{fig7}. The upper panel has $T < 0.1$ 
thermodynamics that is accessible to ED/DMRG. Open points are $T(\alpha,N)$ with $N = 96$ and in some cases also $128$ and $152$. The $S^\prime(T,\alpha)$ 
maxima are solid points $T^*(\alpha)$. Lines at $T < T(\alpha,N)$ or $T^*(\alpha)$, whichever is lower, are Eq.~\ref{eq:apprxen} with exponent $\gamma(\alpha)$ 
in Table~\ref{tab2}. 

$S^\prime(T,\alpha)$ increases continuously to $T^*(\alpha)$ in gapped chains and 
is initially constant in gapless chains. The linear regime between $T^*(\alpha)$ and $T(\alpha,N)$ shrinks to $T^*(\alpha)$ 
with increasing $\alpha$ and $N$, as shown explicitly for $\alpha \ge 0.45$. As best seen for $\alpha = 0.40$ in the lower panel, $T^*(0.4)$ is slightly underestimated
because $S^\prime(T,0.4)$ is not quite constant. The abrupt increase of $S^\prime(T,\alpha)$ to $T^*(\alpha) < 0.01$ is a general result 
for small $\Delta(\alpha)$. The crossing of $C(T,\alpha)/T$ curves with increasing $T$ in the lower panel follows from entropy 
conservation since the area under $S^\prime(T,\alpha)$ is $\ln2$ for any frustration. The area is conserved to better than $1\%$.  

The exponent $\gamma(\alpha)$ in Eq.~\ref{eq:apprxen} increases with $\alpha$ since $S^\prime(T,\alpha)$ and $\Delta(\alpha)/T^*(\alpha)$ 
increase with $\alpha$. The spin gap opens at $\alpha_c = 0.2411$ where $T^*(\alpha_c) = 0$.  Just above $\alpha_c$ we have $T^* = 0+$ and 
slope $A(\alpha_c)$ at $T > T^*$. Eq.~\ref{eq:apprxen} with $\gamma(\alpha) = -1$ and $\Delta(\alpha) \rightarrow 0$ returns linear $S(T,\alpha)$. 
Increasing $\gamma(\alpha)$ for $\alpha \ge \alpha_c$ follows directly from $S(T,\alpha)$ even though the present results are limited to $\alpha \ge 0.35$ 
and Eq.~\ref{eq:apprxen} is phenomenological. 

Spin correlations account qualitatively for increasing $S(T,\alpha)$ with $\alpha$ in gapless chains and 
increasing $S(T^*,\alpha)$ in gapped chains. Separate evaluation of $N^{-1}\ln Q(T,\alpha,N)$ and $E(T,\alpha,N)/T$ indicates 
that the internal energy per site is considerably larger at low $T$. The internal energy density of Eq.~\ref{eq:j1j2} is
\begin{equation}
	E \left (T,\alpha \right ) = {C}_{2} \left (T,1,\alpha \right ) +\alpha {C}_{2} \left (T,2,\alpha \right ).
        \label{eq:intene}
\end{equation}
Taylor expansion about $\alpha = 0$ leads to
\begin{eqnarray}
	E \left (T,\alpha \right ) -E \left (T,0 \right ) &=\alpha {C}_{2} \left (T,2,0 \right ) \qquad \qquad
	\nonumber \\ &  +\alpha {\left (\frac 
	{\partial {C}_{2} \left (T,1,\alpha \right )}  {\partial \alpha} \right )}_{0} +O \left ({\alpha} ^ {2} \right ) . 
        \label{eq:intenetaylor}
\end{eqnarray}
The HAF correlation function $C_2(T,2,0)$ between second neighbors~\cite{sato2005} is $C_2(0,2,0) = 0.18204$ while first-neighbor 
correlation function $C_2(0,1,0) = - 0.44315$ becomes less negative with increasing $\alpha$. Both linear terms in Eq.~\ref{eq:intenetaylor} 
are positive. The $T$ dependence is negligible for $T < 0.1$.

Thermal fluctuations are quantified by $C(T,\alpha)/T$. As seen in Fig.~\ref{fig7}, the density of low-energy correlated 
states increases with frustration $\alpha \le \alpha_c$. Correlated states are shifted out of the gap $\Delta(\alpha)$ for $\alpha \ge \alpha_c$, 
thereby increasing the local density of states. The behavior of correlated states is similar to the single-particle picture, 
at least at the level of thermal averages.

\section{\label{sec4} Critical points}
The entropy provides an independent way of identifying critical points between gapless and gapped quantum phases. 
Linear $S(T,\alpha)$ at low $T$ in gapless phases implies $B(\alpha) = 0$ in Eq.~\ref{eq:linearen} and Table~\ref{tab1} while a gap leads to 
$S^\prime(0,\alpha) = 0$ and exponentially small entropy at $T \ll \Delta(\alpha)$. The evaluation of critical points depends on how 
quantitatively Eq.~\ref{eq:linearen} determines the dashed lines in Fig.~\ref{fig6} at $T < T(\alpha,96)$ or $T(\alpha,152)$ for $\alpha = 0.30$. 
Increasing the system size reduces the extrapolated interval while the coefficients $A(\alpha)$ and $B(\alpha)$ in Table~\ref{tab1} 
reflect the numerical accuracy.

\begin{figure}[t]
         \begin{center} \includegraphics[width=\columnwidth]{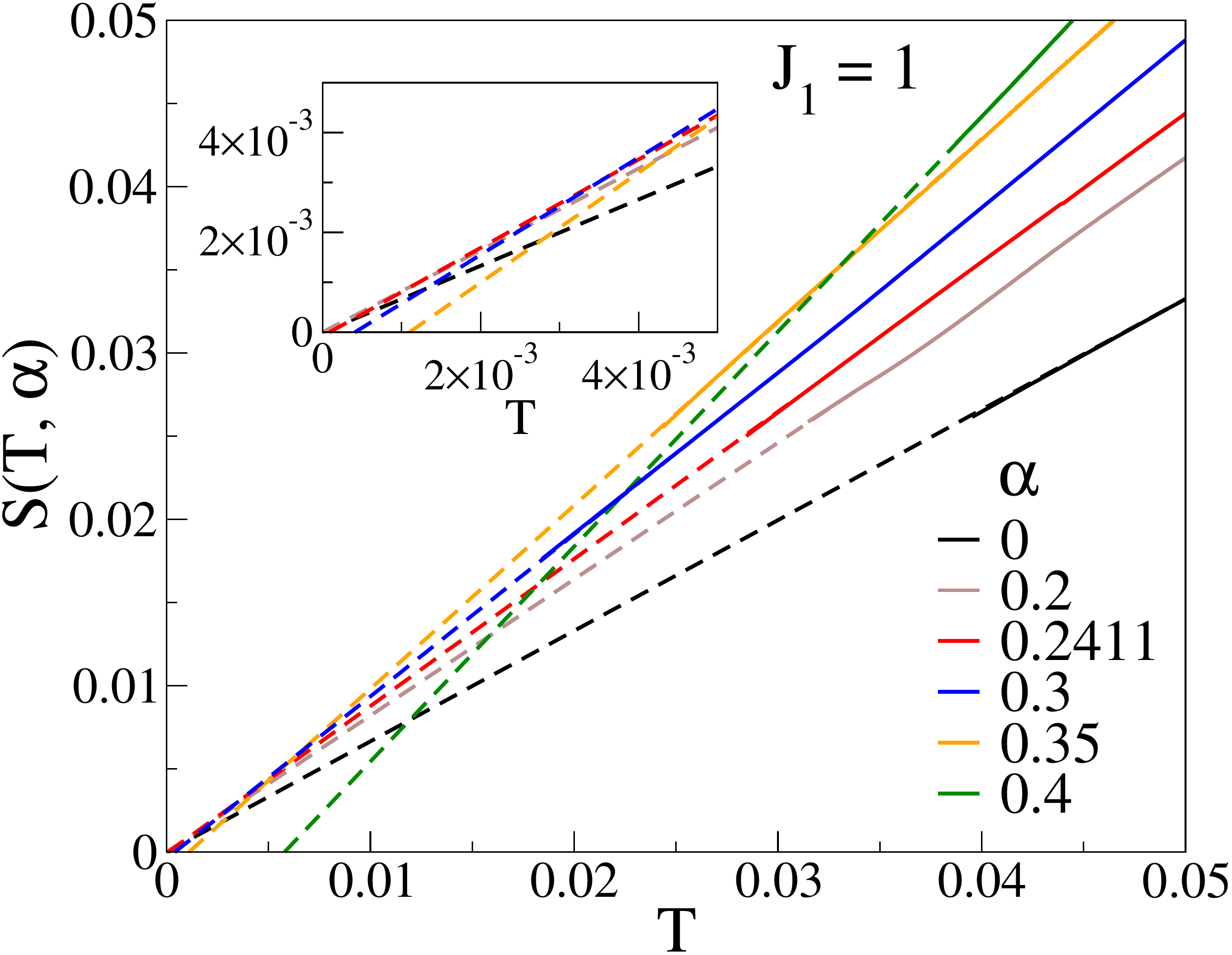}
                 \caption{
$S(T,\alpha)$ to $T = 0.05$ and (inset) $T = 0.005$. Continuous lines are converged for $T > T(\alpha,96)$ or $T(\alpha,152)$ 
		 as noted in the text; dashed lines are linear extrapolation using Eq.~\ref{eq:linearen}.
                 }
 \label{fig8}
 \end{center}
 \end{figure}

Fig.~\ref{fig8} zooms in on $S(T,\alpha)$ up to $T = 0.05 > T(\alpha,96)$ or $T(\alpha,152)$ for $\alpha = 0.30$ where 
continuous lines are converged $S(T,\alpha)$, with $2T/3$ at $\alpha = 0$. The inset magnifies the origin. As noted above, 
frustration initially increases $S(T,\alpha)$. The inset indicates gapped phases at $\alpha = 0.30$ or larger with $B(\alpha) > 0$ 
and a gapless phase at $\alpha = 0.20$. At $\alpha_c = 0.2411$, we find $B(\alpha_c) = 8 \times 10^{-5}$ and consider it to be of zero 
within numerical accuracy. This well-established critical point benchmarks the entropy determination.

The  quantum critical point $\alpha_c = 0.2411$ between the gapless phase and the dimer phase is based on level crossing~\cite{nomura1992} 
and field theory~\cite{haldane82, *haldane82_2nd,kuboki87,affleck88}. As recognized from the beginning, an exponentially small $\Delta(\alpha)$ 
is beyond direct numerical evaluation. However, Okamoto and Nomura~\cite{nomura1992} pointed out that finite systems with nondegenerate 
ground states have a finite-size gap $\Delta^\prime(\alpha,N)$ to the lowest singlet and that gapped phases must have two singlets below 
the triplet. The weak size dependence of the crossing point $\alpha(N)$ at which $\Delta^\prime(\alpha,N) = \Delta(\alpha,N)$ 
yields~\cite{nomura1992} $\alpha_c$ on extrapolating ED results to $N = 24$.

The critical point $1/\alpha_2 = 0.44 \pm 0.01$ ($J_2/J_1 = 2.27 \pm 0.06$) between the gapped incommensurate (IC) and 
gapless decoupled phases is based on level crossing~\cite{mkumar2015} (ED to $N = 28$) and the maximum of the spin structure factor~\cite{soos-jpcm-2016} 
(DMRG to $N = 192$). As mentioned in Section~\ref{sec2}, the chain length is effectively $N/2$ when $J_1$ is small. It is then convenient to work 
with $H(\alpha)/\alpha$ and $J_2 = 1$, $J_1 = 1/\alpha$ in Eq.~\ref{eq:j1j2}. 

\begin{figure}[t]
         \begin{center} \includegraphics[width=\columnwidth]{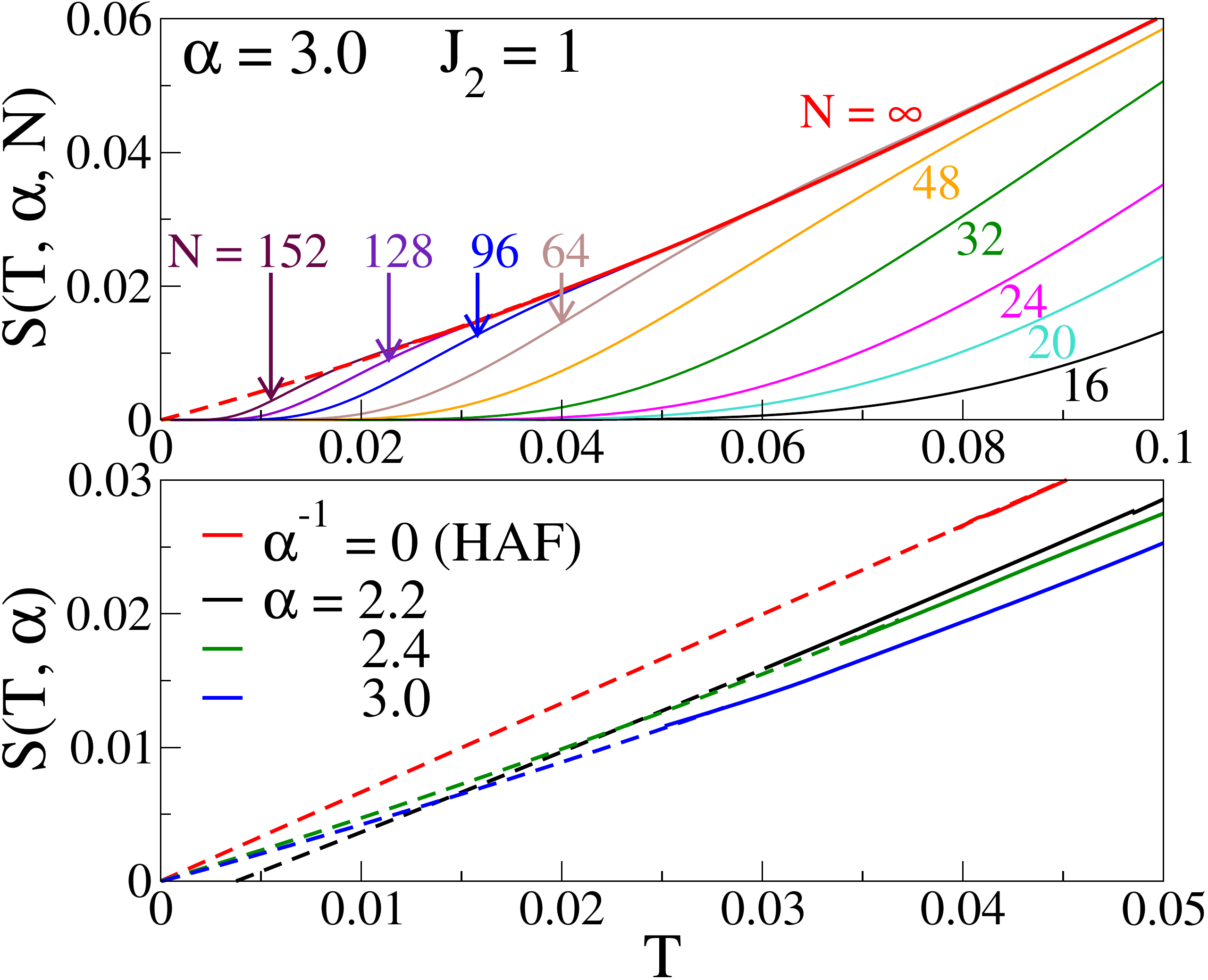}
                 \caption{
Upper panel: Convergence of $S(T,\alpha,N)$ to the thermodynamic limit $S(T,\alpha)$ for $T > T(\alpha,152)$. The dashed line 
		 is extrapolation to $T = 0$. Lower panel: Continuous and dashed lines are converged and extrapolated $S(T,\alpha)$, 
		 respectively, with $N = 96$ for $\alpha = 0$ and 128 for $\alpha = 2.2$ and $2.4$.
		 }
 \label{fig9}
 \end{center}
 \end{figure}

The $\alpha = 3$ entropy $S(T,\alpha,N)$ in Fig.~\ref{fig3} \
is almost equal at low $T$ to $S(T,0,N/2)$. Fig.~\ref{fig9}, upper panel, zooms in on $T \le 0.1$ where convergence to $S(T,\alpha)$ 
holds for $T > T(\alpha,152) = 0.023$. A linear plus quadratic fit to $T = 0.10$ gives the dashed line with $S(0,0.3) = 0$, as does a linear 
fit up to $T = 0.03$. Larger $N$ is more demanding computationally but is needed here since the system is effectively $N/2$. 
The solid and dashed lines in the lower panel are converged and extrapolated $S(T,\alpha)$, respectively, with DMRG to $N = 128$ for $\alpha = 2.4$ 
(gapless) and $2.2$ (gapped). The critical point based on entropy is consistent with other estimates and occurs at 
finite $J_1 = 1/\alpha_2$ rather than at $J_1 = 0$.

\begin{figure}[]
         \begin{center} \includegraphics[width=\columnwidth]{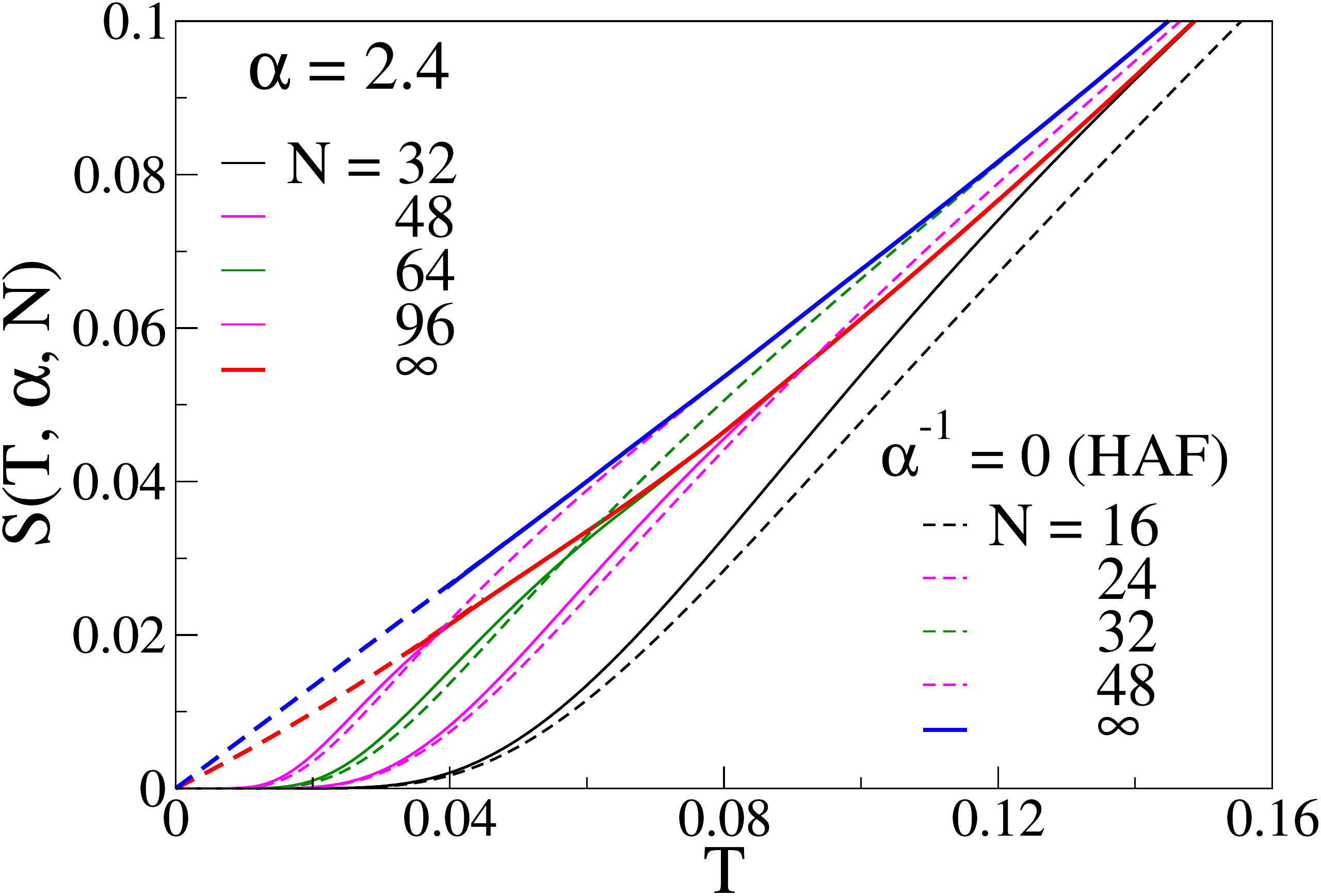}
                 \caption{
$S(T,\alpha,N)$ of Eq.~\ref{eq:j1j2} with $J_2 = 1$, $J_1 = 1/2.4$ (continuous lines) at system size $N$ and with $J_2 = 1$, $J_1 = \alpha^{-1} = 0$ 
		 at $N/2$ (dashed lines). Converged $S(T,\alpha)$ are shown as continuous lines for $T > T(\alpha,N)$ and 
		 extrapolations to $T = 0$ as dashed lines.
		 }
 \label{fig10}
 \end{center}
 \end{figure}

We notice that $S(T,\alpha)$ for $\alpha  > 1$ in Fig.~\ref{fig9}, lower panel, is comparable to or slightly smaller than $S(T,0)$ 
whereas the $\alpha < 1$ entropies in Fig.~\ref{fig7} are considerably larger than the HAF entropy. Even at $\alpha^{-1} = 1/2.4$, 
the $S(T,\alpha,N)$ in Fig.~\ref{fig10} curves are remarkably close to $S(T,0,N/2)$ up to $T = 0.05$; the $\alpha = 3$ curves in Fig.~\ref{fig3} 
are even closer in this interval. The reason is the difference between intrachain spin correlations in Eq.~\ref{eq:intenetaylor} for $\alpha < 1$ 
and spin correlations between sublattices for $\alpha > 1$. With $J_2 = 1$ and $J_1 = \alpha^{-1}$, the Taylor expansion of the 
internal energy about $\alpha^{-1} = 0$ is 
\begin{eqnarray}
	E \left (T, {\alpha} ^ {-1} \right )  -E \left (T,0 \right ) = {\alpha} ^ {-1} {C}_{2} \left (T,1,0 \right ) \nonumber \\ \qquad \quad  + {\alpha} ^ {-1} {\left (\frac {\partial {C}_{2} 
	\left (T,2, {\alpha} ^ {-1} \right )}  {\partial {\alpha} ^ {-1}} \right )}_{0}  +O \left ({\alpha} ^ {-2} \right ) .
        \label{eq:intenetaylorlarge}
\end{eqnarray}
Since $\alpha^{-1} = 0$ corresponds to noninteracting HAFs on sublattices, $C_2(T,1,0) = 0$ and $C_2(T,2,0)$ is the 
first neighbor correlation within sublattices. It has a minimum at $\alpha^{-1} = 0$ and becomes less negative for 
either sign of $J_1$. There is rigorously no $\alpha^{-1}$ term. 

Bond-bond correlation functions provide additional characterization of critical points. The largest separation between bonds (1,2) and ($2r + 1$, $2r + 2$) 
in a chain of $N = 4n$ spins with periodic boundary conditions is at $r = n$. We define the four-spin correlation function at 
frustration $\alpha$ as the ground state expectation value
\begin{equation}
 C_4(2n,\alpha)=\langle G(4n,\alpha) \vert S^z_1 S^z_2 S^z_{2n+1} S^z_{2n+2} \vert G(4n,\alpha) \rangle.
\label{eq:4spincor1}
\end{equation}
Bonds (1,2) and ($2n + 1$, $2n + 2$) are in the same Kekul\'e VB diagram, either $\vert K1 \rangle$ or $\vert K2 \rangle$. 
The next most distant bonds have $2n \rightarrow 2n \pm 1$ in Eq.~\ref{eq:4spincor1} and one bond in $\vert K1 \rangle$, 
the other in $\vert K2 \rangle$. The difference between most and next most distant correlation functions is
\begin{equation}
\qquad D_4(2n,\alpha)=C_4(2n,\alpha)-C_4(2n-1,\alpha).
\label{eq:4spincor_diff}
\end{equation}
Finite $D_4(2n,\alpha) > 0$ as $n \rightarrow \infty$ indicates long-range bond-bond correlations. The correlation functions 
are readily evaluated at the MG point where $D_4(1/2) = 1/32$ for distant bonds. Except for nearby neighbors, bonds in different 
diagrams are uncorrelated, with $C_4(2n - 1,1/2) = 0$, while $C_4(2n,1/2) = 1/16$ for the diagram with both bonds 
and zero for the other diagram.

\begin{figure}[]
         \begin{center} \includegraphics[width=\columnwidth]{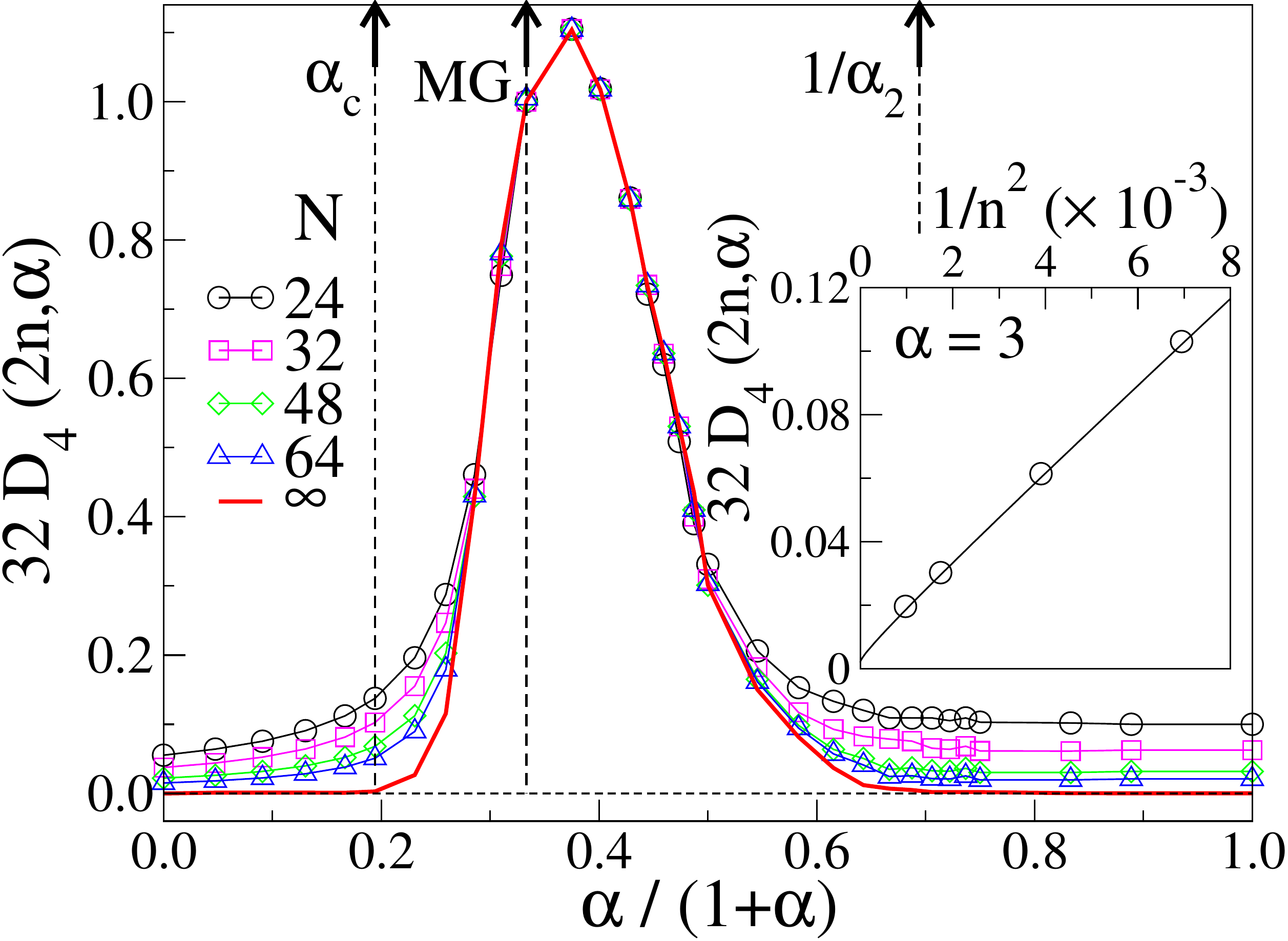}
                 \caption{
			 Ground state bond-bond correlations $D_4(2n,\alpha)$ in Eq.~\ref{eq:4spincor_diff} for $N = 4n$ 
			 spins and frustration $\alpha$ in Eq.~\ref{eq:j1j2}. Finite $D_4(2n,\alpha)$ indicates that different 
			 correlations between the most and next most distant bonds persist in the thermodynamic limit. Arrows 
			 mark the critical points and MG point. Inset: Two-spin correlation functions within sublattices, Eq.~\ref{eq:4spincor_diff_decoupled}, 
			 that go as $n^{-2}$ in the decoupled phase, $1/\alpha < 1/\alpha_2$.
		 }
 \label{fig11}
 \end{center}
 \end{figure}

Fig.~\ref{fig11} shows bond-bond correlations $D_4(2n,\alpha)$ in systems of $N = 4n$ spins over the entire range from $J_2 = 0$ to $J_1 = 0$. 
The red line is based on $1/N$ extrapolations of $D_4(2n,\alpha)$. The gapped phases between $\alpha_c$ and $1/\alpha_2$ have long-range 
bond-bond correlations that exceed unity at $\alpha = 0.60$. The spin gap opens quite differently with increasing $\alpha < 1$ and increasing $\alpha^{-1} < 1$. 
The structure factor peak~\cite{soos-jpcm-2016} is finite at wavevector $q = \pi$ in the dimer phase $\alpha_c \le \alpha \le 1/2$. 
The peaks are finite at $\pi \pm q(\alpha)$ in the IC phase with $q(\alpha) = 0$ at $\alpha = 1/2$ and increasing to $\pi/2$ at $\alpha_2$. 
The gapless phase at small $\alpha$ has quasi-long-range spin correlations $C_2(p,\alpha)$ while the gapless phase at large $\alpha$ has 
quasi-long-range $C_2(2p,\alpha)$ within sublattices.

Spins in different sublattices are uncorrelated when $J_1 = 0$ ($\alpha \rightarrow \infty$). 
The four-spin correlation functions in Eq.~\ref{eq:4spincor_diff} then reduce to two-spin correlations within sublattices 
\begin{eqnarray}
	D_4(2n,\alpha \rightarrow \infty) &=  {C}_{2} \left (2n,0 \right ) \times \qquad \qquad \qquad \qquad  \nonumber \\  
	& \left ({C}_{2} \left (2n,0 \right ) - {C}_{2} \left (2n-2,0 \right ) \right ) .
\label{eq:4spincor_diff_decoupled}
\end{eqnarray}
Since the sublattice HAF correlations go as $(-1)^n/2n$, $D_4(2n,\alpha)$ decreases as $1/n^2$ when $J_1 = 0$. That is indeed 
the case in Fig.~\ref{fig11} as shown in the inset for $\alpha = 3$. The weak dependence on $\alpha > 2.4$ is additional 
evidence that sublattice spin correlations are hardly sensitive to $J_1$. On the contrary, $\alpha < 1$ correlations are 
very sensitive to $J_2$ since the second neighbor $C_2(2,0) > 0$ changes sign at $\alpha = 1/2$. 

The expansion of the ground state $\vert G(4n) \rangle$ in the correlated real-space basis of $N$-spin VB diagrams 
is well defined~\cite{ramasesha84} for arbitrarily large $N = 4n$. The dimension of the singlet sector is
\begin{equation}
R \left (4n \right ) =\frac {\left (4n \right ) !}  {\left (2n \right ) ! \left (2n+1 \right ) !}.
\label{eq:vb}
\end{equation}
The Kekul\'e diagrams $\vert K1 \rangle$ and $\vert K2 \rangle$ are the only ones long-range bond-bond order in arbitrarily 
large systems. Accordingly, their expansion coefficients are macroscopic in the thermodynamic limit of gapped $J_1-J_2$ models with finite $\Delta(\alpha)$, 
doubly degenerate ground state and $D_4(2n,\alpha) > 0$ as $n \rightarrow \infty$. 

\section{\label{sec5} Magnetic susceptibility}
\begin{figure}[]
         \begin{center} \includegraphics[width=\columnwidth]{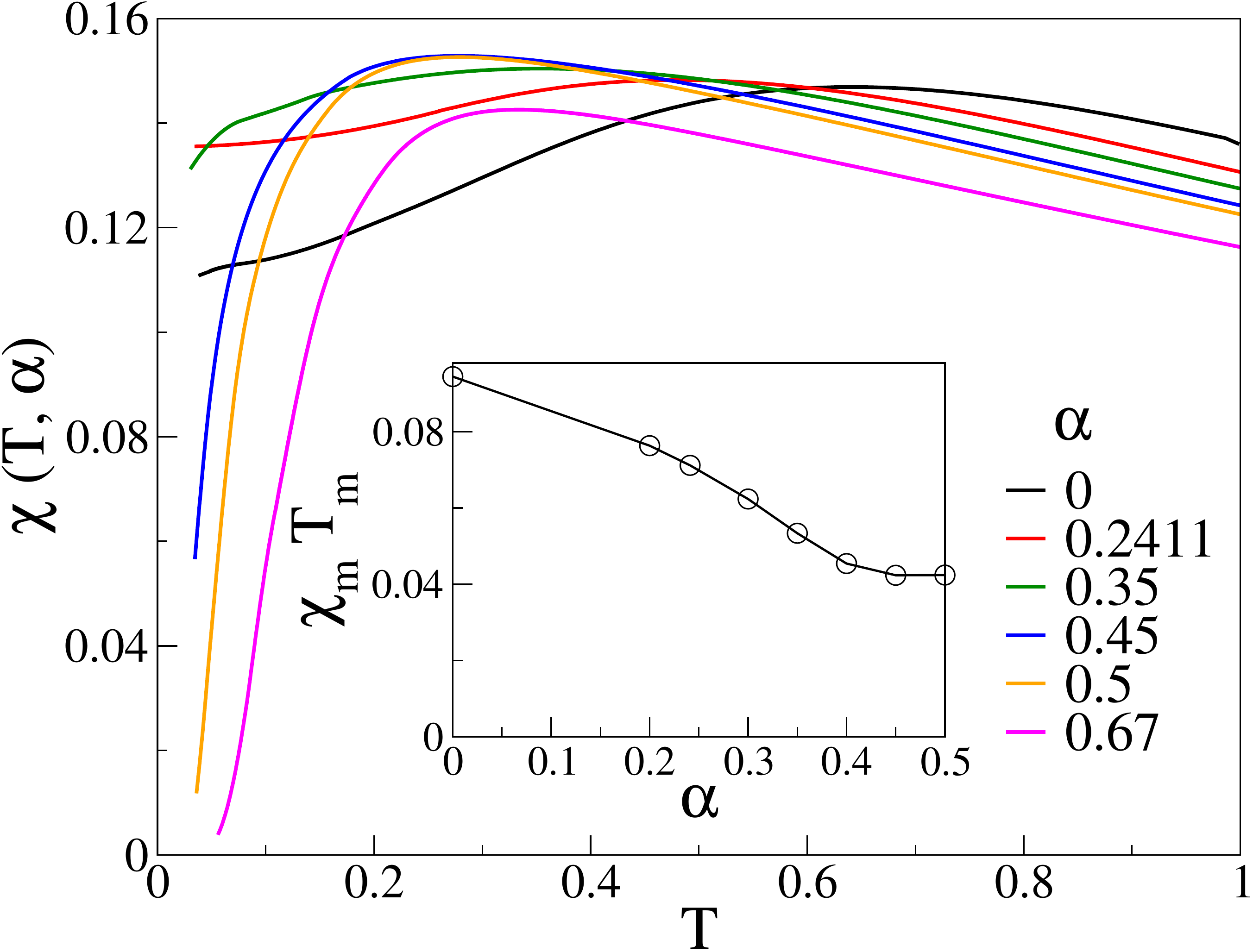}
                 \caption{
Converged $\chi(T,\alpha)$ at $T > T(\alpha,N)$ with $N = 96$ for $\alpha \le 0.35$
and $N = 152$ for $\alpha \ge 0.45$. 
		 Inset: the product $\chi_m T_m$ at the maximum specifies $\alpha$.} 
 \label{fig12}
 \end{center}
 \end{figure}

Crystallographic data specifies the unit cells of materials with strong exchange within chains or layers. The measured 
molar magnetic susceptibility $\chi(T)$ of chains with one spin-1/2 per unit cell can be compared the $\chi(T,\alpha)$ of 
1D models such as $H(\alpha)$ in Eq.~\ref{eq:j1j2}. Long ago, Bonner and Fisher~\cite{bonner64} used ED to $N = 12$, 
insightful extrapolations and the $T = 0$ result to obtain converged $\chi(T,0)$ for $T/J_1 > 0.1$ and a good approximation 
for the HAF down to $T = 0$. Now ED to $N = 24$ yields converged $\chi(T,\alpha)$ and $C(T,\alpha)$ to lower $T$ and DMRG for $N > 24$ 
extends the range to $T > T(\alpha,N)$ in spin-1/2 chains with isotropic exchange. Susceptibility data on many materials, 
both inorganic and organic, are consistent with HAFs. Physical realizations are quasi-1D due to other interactions such 
magnetic dipole-dipole interactions or exchange between spins in different chains.

Fig.~\ref{fig12} shows converged $\chi(T,\alpha)$ for $T > T(\alpha,96)$ for $\alpha \le 0.35$ and $T(\alpha,152)$ for $\alpha \ge 0.45$. 
The increase with $\alpha$ at low $T$ for small or no gap is similar that of $C(T,\alpha)/T$ in Fig.~\ref{fig4}. We again find 
quantitative agreement for $T > 0.2$ with previous $\chi(T,\alpha)$ results~\cite{okunishiprb,white2005}. The maximum $\chi_m(T,\alpha)$ at $T_m(\alpha)$ 
shifts to lower $T$ in both gapless and gapped chains up to $\alpha = 0.45$. The product $\chi_m T_m$ in the inset specifies $\alpha$. 
Converged $\chi(T,\alpha)$ for $T > T(\alpha,152)$ is almost quantitative at $\alpha = 0.67$ or 0.50.

The Peierls instability applies to spin-1/2 chains with linear spin-phonon coupling in the $\alpha < 1$ sector of Eq.~\ref{eq:j1j2}. 
The spin-Peierls transition at $T_{SP}$ leads at lower $T$ to a dimerized chain with two spins per unit cell, provided that 
competing 2D or 3D interactions do not induce other transitions. Susceptibility data to 950 K fixed~\cite{fabricius1998} $\alpha = 0.35$ 
in the inorganic spin-Peierls (SP) crystal CuGeO$_3$ with $J_1 = 160$ K and $T_{SP} = 14$ K. Data to 350 K fixed~\cite{jacob1976} $\alpha = 0$ 
in an organic SP crystal with $J_1 = 79$ K and $T_{SP} = 12$ K. We have recently modeled~\cite{sudipsp2020} both SP transitions successfully 
using correlated states both below and above $T_{SP}$. The analysis of high $T$ data is primarily a matter of identifying the proper 
model, the appropriate version of Eq.~\ref{eq:j1j2}, bearing in mind that isotropic exchange (no spin-orbit coupling) is an 
approximation for spins centered at metallic ions.

Extrapolation is required to obtain converged $\chi(T,\alpha)$ in the interval $0 \le T < T(\alpha,96)$ or $T(\alpha,152)$. 
There are three cases: (1) In gapless chains, the weak $T$ dependence of $\chi(T,\alpha)$ is readily extrapolated to finite $\chi(0,\alpha)$. 
In gapped chains, we distinguish below between (2) $T(\alpha,N) < T^*(\alpha)$ and (3) $T^*(\alpha) < T(\alpha,N)$ as 
discussed for $S(T,\alpha)$. We note that the ground-state degeneracy in Fig.~\ref{fig1} leading to $S(0,\alpha) = N^{-1} \ln 2$ in gapped 
chains is readily seen when $N$ exceeds the gap $\Delta^\prime(\alpha,N)$ between the ground state and the lowest singlet excited state. 
The zero-point extropy of finite chains interferes with convergence to the thermodynamic limit. Convergence to $\chi(T,\alpha)$ is simpler in this 
respect and is achieved at system size $N \sim 100$ in $J_1-J_2$ models with large $\Delta(\alpha)$.

\begin{figure}[]
         \begin{center} \includegraphics[width=\columnwidth]{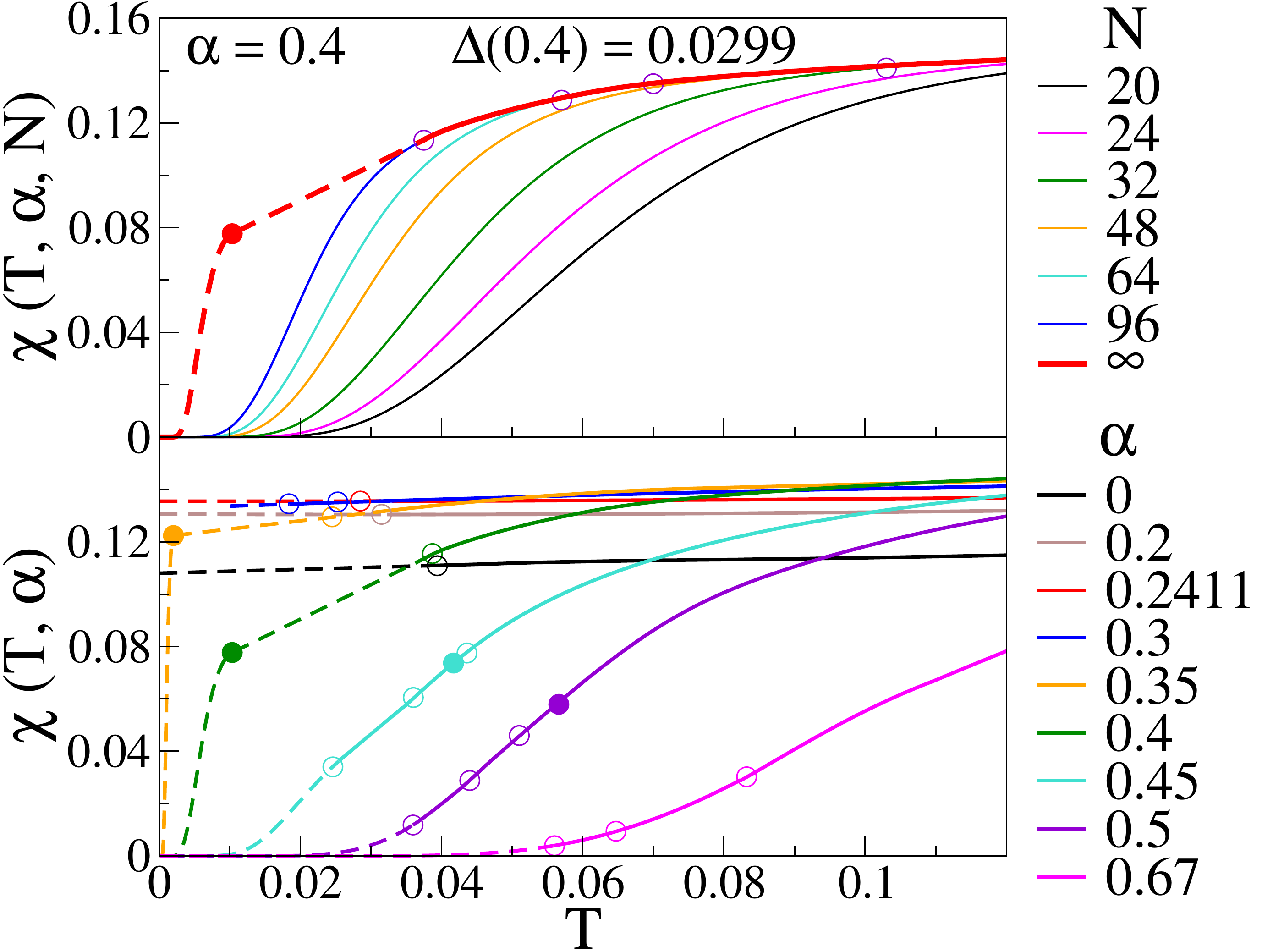}
                 \caption{
Upper panel: Convergence of $\chi(T,0.4,N)$ with system size $N$ to $S(T,0.4)$ for $T > 0.04$ indicated by open points. 
		 The dashed red line is Eq.~\ref{eq:apprxchi} for $T < T^*(0.4)$, the solid point, and for the $T > T^*(0.4)$ 
		 extrapolation discussed in the text. Lower panel: Continuous lines are converged $\chi(T,\alpha)$ at $T > T(\alpha,N)$; 
		 dashed lines at lower $T$ are discussed in the text. Open points are $T(\alpha,96)$ for all $\alpha$, $T(\alpha,152)$ for $\alpha = 0.30$ and $\alpha,N$ 
with $N = 128$ and $152$ for $\alpha \ge 0.045$. Solid points are $T^*(\alpha)$ for gapped chains.} 
		  
 \label{fig13}
 \end{center}
 \end{figure}

For gapped chains, we took the functional form for $S(T,\alpha)$ in Eq.~\ref{eq:apprxen} aside from the exponent $\eta(\alpha)$ 
\begin{equation}
        \chi(T,\alpha)=c(\alpha)T^{-\eta(\alpha)}\exp(-\Delta(\alpha)/T).
        \label{eq:apprxchi}
\end{equation}
The range is again $T = 0$ to the lower of $T^*(\alpha)$ or $T(\alpha,N)$. We start with $T(\alpha,N) < T^*(\alpha)$.
As seen in Fig.~\ref{fig2}, $\chi(T,0.5,152)$ is close to convergence and the larger gap at $\alpha = 0.67$ ensures even faster 
convergence. Convergence at $N = 152$ in Fig.~\ref{fig13} reaches $T(\alpha,152)$ and small $\chi(T,\alpha)$. 
We determine $\eta(\alpha)$ for $\alpha \ge 0.45$ by a least squares fit of Eq.~\ref{eq:apprxchi} to $\chi(T,\alpha,152)$ 
up to $T(\alpha,152)$. The dashed line in Fig.~\ref{fig2} has $\eta(0.5) = 1.23$ for $T < 0.0246$.

When $T^*(\alpha) < T(\alpha,96)$, we rely on both Eq.~\ref{eq:apprxchi} and extrapolation. Fig.~\ref{fig13}, upper panel, 
shows convergence with size at $\alpha = 0.40$. Open points are decreasing $T(0.4,N)$ with increasing $N$. The solid point is $T^*(0.4) = 0.0103$ 
in Table~\ref{tab2} based on the entropy. We extrapolate converged $\chi(T,\alpha)$ from $T(\alpha,96)$ to $T^*(\alpha)$ as 
$A^\prime(\alpha)T - B^\prime(\alpha)$ and match the magnitude and slope of Eq.~\ref{eq:apprxchi} to evaluate $\eta(\alpha)$. We obtain
\begin{equation}
\eta \left (\alpha \right ) =\frac {\Delta \left (\alpha \right )} {{T} ^ {*} (\alpha)} - \frac {1} 
	{1- {{B} ^ {\prime} \left (\alpha \right )} / {{A} ^ {\prime} \left (\alpha \right ) {T} ^ {*} \left (\alpha \right )}} .   
        \label{eq:eta}
\end{equation}
The exponents $\eta(\alpha)$ in Table~\ref{tab2} are based on Eq.~\ref{eq:eta} for $\alpha \le 0.40$ and least squares fits for $\alpha \ge 0.45$. 

The lower panel of Fig.~\ref{fig13} shows converged $\chi(T,\alpha)$ of gapless chains with finite $\chi(0,\alpha)$
and gapped chains. Open points are $T(\alpha,96)$ for all $\alpha$, $T(\alpha,152)$ for
$\alpha = 0.30$ and $T(\alpha,N)$ at $N = 128$ and $152$ for $\alpha \ge 0.45$. 
Solid points are $T^*(\alpha)$, the $S^\prime(T,\alpha)$ maxima. Once again, modeling the small gap at $\alpha = 0.30$ and $T^*(0.3) < 10^{-3}$ 
requires considerable larger systems.

Converged $\chi(T,\alpha)$ in gapped chains at $T < T^*(\alpha)$ indicates power-law deviations with exponents $\eta(\alpha)$ 
in Eq.~\ref{eq:apprxchi} and Table~\ref{tab2}. We find that $\eta(\alpha)$ is almost constant up to $\alpha = 0.45$ and then 
decreases significantly at $\alpha = 0.50$ and 0.67. The $\chi(T,\alpha)$ knee at $T^*(\alpha)$ in Fig.~\ref{fig13} for $\alpha = 0.35$ 
or 0.40 requires $\eta(\alpha) \sim 2.7$. There is no knee at $\alpha = 0.50$ or 0.67 with $\eta(\alpha) < 2$. 
We speculate that $\eta(0.45) = 2.8$ is due to the steep slope at $T^*(0.45) = 0.042$.

Both $S(T,\alpha)$ and $\chi(T,\alpha)$ become exponentially small in gapped chains as $T \rightarrow 0$, with exponents $\gamma(\alpha)$ 
and $\eta(\alpha)$ that describe the thermal and magnetic fluctuations, respectively. To focus on 
deviations from $\exp(-\Delta(\alpha)/T)$, we consider the ratio
\begin{equation}
R \left (T,\alpha \right ) =\frac {S(T,\alpha)} {4T \chi \left (T,\alpha \right )} \equiv \frac {S(T,\alpha)}  {\rho(T,\alpha)} .
	\label{eq:ratio}
\end{equation}
Since the high $T$ limit of $\chi(T,\alpha)$ is the Curie law, $1/4T$ in reduced units, the spin density $\rho(T,\alpha)$ 
defined in Eq.~\ref{eq:ratio} is unity in that limit; $\rho(T,\alpha)$ is the effective density of free spins at temperature $T$. 
The high-$T$ limit is $R(T,\alpha) = \ln 2$ since $S(T,\alpha)$ goes to $\ln 2$, independent of $\alpha$ or $N$. The ratio 
quantifies the relative magnitudes of thermal and magnetic fluctuations.

\begin{figure}[]
         \begin{center} \includegraphics[width=\columnwidth]{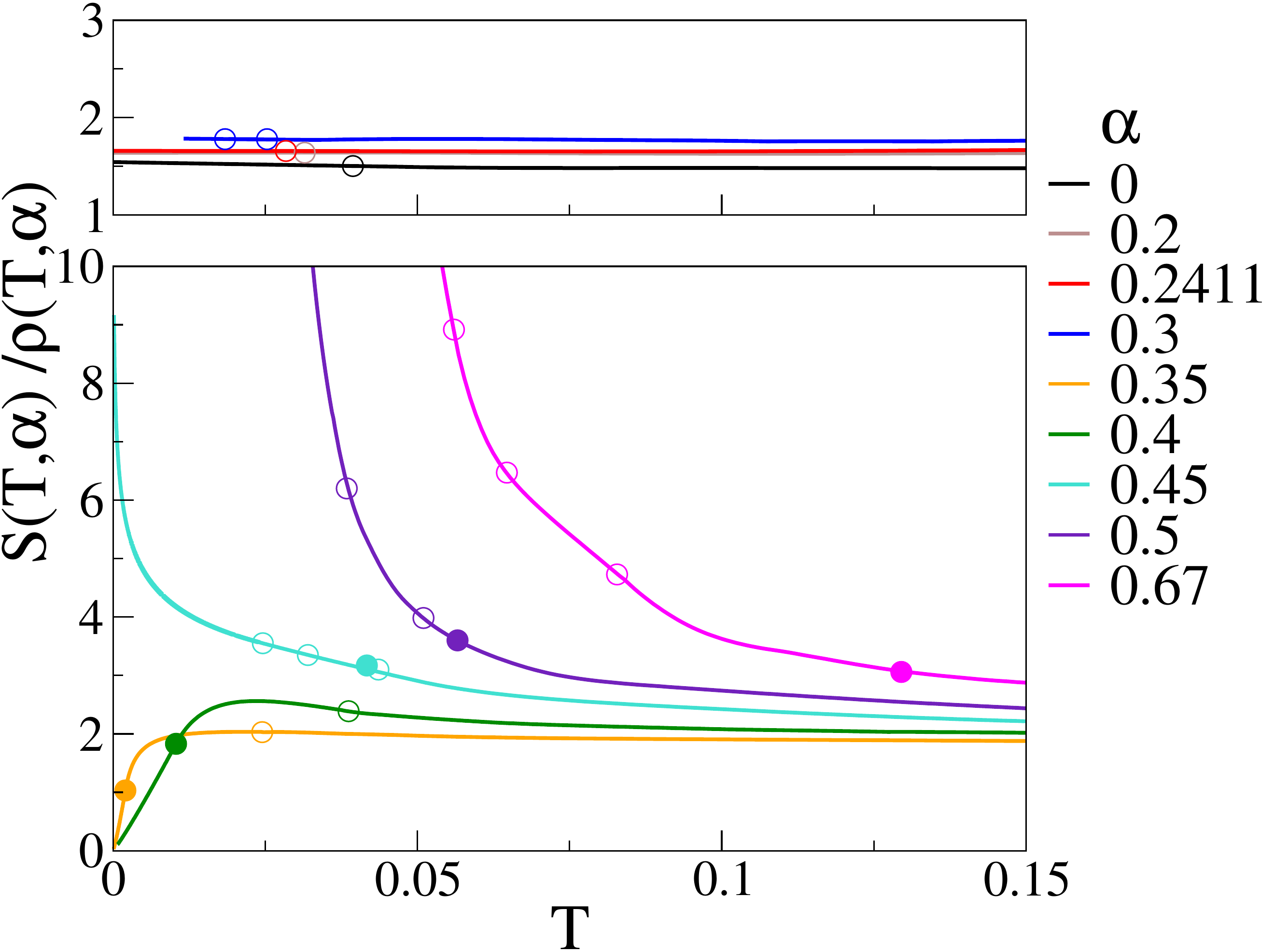}
                 \caption{
Ratio $S(T,\alpha)/\rho(T,\alpha)$ at $\alpha \le 0.30$ (upper panel) and $\alpha \ge 0.35$ (lower panel). The ratio is constant for $\alpha \le 0.30$. 
		 Converged results for $T > T(\alpha,N = 96)$, shown as open points, are extrapolated to $T = 0$ for gapless chains; $N = 96$ 
		 and 152 points are shown at $\alpha = 0.30$. The $T$ dependent ratios in the lower panel have $T(\alpha,N)$ at $N = 96$ 
		 for all $\alpha$ and also $N = 128$ and 152 for $\alpha \ge 0.45$. Solid points are $T^*(\alpha)$, the maxima of $S^\prime(T,\alpha)$.
		 }
 \label{fig14}
 \end{center}
 \end{figure}

Fig.~\ref{fig14}, upper panel, shows $R(T,\alpha)$ up to $T = 0.15$, with open points at $T(\alpha,96)$ for all $\alpha$ and at $T(\alpha,152)$ for $\alpha = 0.30$. 
The ratio is almost constant for gapless chains and for $\alpha = 0.30$. Except for $\alpha = 0.30$, we extrapolate to $T = 0$ and find $R(0,0) = 1.54$, 
within $7\%$ of the exact $\pi^2/6$ for the HAF. The difference is mainly due to logarithmic corrections that, as shown in Fig.~\ref{fig1} of Ref.~\onlinecite{johnstonprl}, 
increase $\chi(T,0)$ by almost $6\%$ at $T = 10^{-4}$ and by more than $6\%$ at $T = 10^{-3}$. Such corrections and rigorous $T \rightarrow 0$ 
limits are beyond the ED/DMRG method. Frustration slightly increases $R(T,\alpha)$ in gapless chains; $S(T,\alpha)/T$ evidently increases 
faster than $\chi(T,\alpha)$. Constant $R(T,\alpha)$ in gapless chains follows from the $S^\prime(T,\alpha)$ and $\chi(T,\alpha)$ 
results in Figs.~\ref{fig7} and ~\ref{fig13}.

The remarkable dependence of $R(T,\alpha)$ on frustration in gapped chains is seen in Fig.~\ref{fig14}, bottom panel. Solid points are $T^*(\alpha)$. 
Open points at $\alpha = 0.35$ and 0.40 are $T(\alpha,N)$ for $N = 96$, and at $\alpha \ge 0.45$ are $T(\alpha,N)$ for $N = 96$, 128 and 152. 
Converged $S(T,\alpha)$ and $\chi(T,\alpha)$ give $R(T,\alpha)$ for $T > T(\alpha,N)$, the largest system studied. The exponents $\gamma(\alpha)$ and $\eta(\alpha)$ in Table~\ref{tab2} govern the $T$ 
dependence at $T < T(\alpha,N)$. Within this approximation, $R(T,\alpha)$ is proportional to $T^{\eta-\gamma-1}$. We have $R(0,\alpha) = 0$ 
when $\eta(\alpha) > \gamma(\alpha) + 1$, divergent $R(0,\alpha)$ when $\eta(\alpha) < \gamma(\alpha) + 1$ and constant $R(T,\alpha)$ up to $T(\alpha,N)$ 
when $\eta(\alpha) = \gamma(\alpha) + 1$. The weak $T$ dependence at intermediate $\alpha = 0.45$ is nominally $T^{-0.18}$. The spread between $R(T,0)$ and $R(T,0.67)$ decreases at higher $T$: from 0.882 to 0.999 at $T = 2$ and from 0.753 to 0.792 at $T = 6$. 
The high $T$ limit is $R(T,\alpha) = \ln 2$. 

The $T \rightarrow 0$ limit of $R(T,\alpha)$ depends on the phenomenological Eqs.~\ref{eq:apprxen} and ~\ref{eq:apprxchi}. But the intermediate 
nature of $\alpha = 0.45$ in Fig.~\ref{fig14} is evident for converged $R(T,\alpha)$, as is the strong dependence on frustration up to $T = 0.15$. The entire $\alpha = 0.5$ curve shown is converged, with $R = 12.4$ at $T(0.5,152)=0.031$. We suggest a qualitative interpretation in terms of $\Delta(\alpha)$. The $\alpha$
dependence of $R(T,\alpha)$ decreases when $T > \Delta(\alpha)$ 
and disappears at high $T$ as noted above. Almost constant $R(T,\alpha)$ for $T > \Delta(\alpha)$ requires $T > 0.03$ for $\alpha \le 0.40$, $T > 0.11$ 
for $\alpha = 0.45$ and $T > 0.23$ or $0.43$ for $\alpha = 0.50$ or 0.67. The internal energy contribution to $S(T,\alpha)$ in the numerator starts 
as $E(T,\alpha) = 3\beta \Delta(\alpha) \exp(-\beta\Delta(\alpha))$ in gapped chains while $\rho(T,\alpha)$ in the denominator starts as 
$S(S + 1)\exp(-\beta\Delta(\alpha))$ with $S = 1$ for a triplet. Then $\Delta(\alpha)/T < 1$ leads to the weak T 
dependence of $R(T,\alpha)$ found for $\alpha \le 0.40$ while $\Delta(\alpha)/T > 1$ rationalizes the strong $T$ dependence for $\alpha = 0.50$ or 0.67.

%%%%%%%%%%%%%%%%%%%%%%%%%%%%%%%%%%%%%%%%%%%%%%%%%%%%%%%%%%%%%%%%%%%%%%%%%%%%%

\section{\label{sec6} Discussion}
We have obtained the low $T$ thermodynamics of the antiferromagnetic $J_1-J_2$ model, Eq.~\ref{eq:j1j2},  with variable frustration $\alpha$ 
in both the $\alpha < 1$ and $\alpha > 1$ sectors. The thermodynamics of strongly correlated models are largely unexplored unless the Bethe ansatz 
is applicable. Considerably more is known about the quantum ($T = 0$) phases of correlated 1-D spin chains. The ground-state degeneracy, elementary 
excitations and critical points provide important guidance for thermodynamics. It is advantageous to perform DMRG at both $T = 0$ and finite $T$. 
The principal difference is that hundreds of low-energy states are targeted at finite $T$ at each system size instead of the ground state.

We compared ED/DMRG results with previously reported thermodynamics~\cite{okunishiprb,white2005} down to $T = 0.05$ and found quantitative agreement at $T > 0.2$, good agreement down to $T \sim 0.1$ 
and limited agreement at lower $T$. Thermodynamics down to $T \sim 0.01$ is demonstrated for the entropy $S(T,\alpha)$, spin specific heat $C(T,\alpha)$ and 
magnetic susceptibility $\chi(T,\alpha)$ by following the size dependence and extrapolation. Larger $N$ is accessible if needed, but the $T \rightarrow 0$ 
limit always requires extrapolation. DMRG to system size $N = 96$, and occasionally $N = 128$ or 152, yields converged $S(T,\alpha)$ or $\chi(T,\alpha)$ 
down to $T(\alpha,N) < 0.05$ in Table~\ref{tab1} before any extrapolation. The main results are converged low $T$ thermodynamics of the antiferromagnetic $J_1-J_2$ 
model over the entire range of frustration $\alpha < 1$ within a chain and frustration $\alpha^{-1} < 1$ between HAFs on sublattices.

We note that the entropy has received far less attention than the magnetic susceptibility or the spin specific heat. To be sure, $\chi(T)$ and $C(T)$ 
are directly related to experiment. But the mathematical physics of the models themselves is the primary motivation for theoretical and 
computational studies of quantum phases, symmetries and excitations. The size dependence of $S(T,\alpha,N)$ yields converged $S(T,\alpha)$ 
that we have exploited in this paper. The $T$ 
dependence provides an independent way of finding and evaluating quantum critical points. Additional evidence for $\alpha_2 = 2.27 \pm 0.06$ 
was an initial motivation. We also studied the difference between frustrating second-neighbor exchange $\alpha < 1$ in a chain with $J_1 = 1$ 
and frustrating exchange $\alpha^{-1}$ between HAFs with $J_2 = 1$ on sublattices of odd and even-numbered sites. Long-range bond-bond 
correlations in gapped phases illustrate other differences.

Converged $S(T,\alpha)$ directly show the $S^\prime(T,\alpha)$ maxima $T^*(\alpha)$ in Table~\ref{tab2} of $J_1-J_2$ models with $\alpha \ge 0.45$. 
Extrapolation and the phenomenological Eq.~\ref{eq:apprxen} lead to $T^*(\alpha)$ in chains with smaller $\Delta(\alpha)$. The power 
law $T^{-\gamma(\alpha)}$ modifies the $\exp(-\Delta(\alpha)/T)$ dependence on the spin gap $\Delta(\alpha)$. The exponent $\gamma(\alpha)$ 
in Table~\ref{tab2} increases with frustration. Fig.~\ref{fig7} 
shows $S^\prime(T,\alpha) = C(T,\alpha)/T$ and the shifting of correlated states shift out of the gap with increasing $\alpha$. Converged $\chi(T,\alpha)$ for $T > T(\alpha,96)$ in Fig.~\ref{fig13} clearly distinguishes between gapless chains with finite $\chi(0,\alpha)$ and gapped chains with the $T^{-\eta(\alpha)}$ factor in Eq.~\ref{eq:apprxchi} for $T \le T^*(\alpha)$.   

The ratio $R(T,\alpha) = S(T,\alpha)/4T\chi(T,\alpha)$ in Fig.~\ref{fig14} 
compares thermal and magnetic fluctuations. It is almost constant in gapless chains up to $T = 0.15$ and increases slightly with $\alpha$. 
In gapped chains, $R(T,\alpha)$ highlights the exponents $\gamma(\alpha)$ and $\eta(\alpha)$ since the spin gap divides out. The ratio 
decreases strongly with increasing $T$ for large gaps $\alpha > 0.45$ but increases with $T$ for $\alpha < 0.45$. 

%ED/DMRG is a general approach to the thermodynamics of correlated 1D models. Larger systems can be studied 
%with additional effort when called for, as illustrated above for both $\alpha < 1$ and $\alpha > 1$. Application to the $T < 0.10$ 
%thermodynamics of the $J_1-J_2$ model with $\alpha < 1$ and $\alpha > 1$ in Eq.~\ref{eq:j1j2} generates quantitative numerical results that, 
%in turn, should lead to better understanding of correlated spin states. 

ED/DMRG is a general approach to the thermodynamics of correlated 1D models. Spin-Peierls 
systems have chains with two spins per unit cell and gap $\Delta(T)$ for $T < T_{SP}$.
The gap increases on cooling and suppresses correlations between spin separated by 
more than $1/\Delta(T)$. The method then holds down to $T = 0$ and has successfully modeled~\cite{sudipsp2020} the two best characterized SP 
systems. The restriction to 1D can be relaxed slightly. Quasi-1D materials with small interchain $J^\prime$ 
compared to intrachain $J$ have long been modeled using the random-phase approximation~\cite{shin_trellis98}. The 1D susceptibility $\chi(T)$ 
is modified as $1/[1 + A(J^\prime/J)\chi(T)]$, where A depends on the model.
Thermodynamics at $T < 0.10$ make it possible to resolve corrections to isotropic exchange due to spin-orbit coupling or other 
small magnetic interactions. The ED/DMRG returns the thermodynamics of the $J_1 - J_2$ model down to $T/J_1 \sim 0.01$ 
for $\alpha < 1$ or $T/J_2 \sim 0.01$ for $1/\alpha < 1$. Quantitative numerical should in turn lead to better understanding of correlated spin states.

%%%%%%%%%%%%%%%%%%%%

\begin{acknowledgments}
MK thanks SERB for financial support through grant sanction number CRG/2020/000754.
SKS thanks DST-INSPIRE for financial support.
\end{acknowledgments}

%%%%%%%%%%%%%%%%%%%%%%%%%%%%%%%%%%%%%%%%%%%%%%%%%%%%%%%%%%%%%%%%%%
%																 %
%																 %
%		REFERENCES											     %
%																 %
%																 %
%%%%%%%%%%%%%%%%%%%%%%%%%%%%%%%%%%%%%%%%%%%%%%%%%%%%%%%%%%%%%%%%%%


%\bibliography{ref_thermo6}
\end{document}